\documentclass{aa}  

\usepackage{graphicx}
\usepackage{lscape}
\usepackage{amsmath}
\usepackage{todonotes}
\begin{document}
   \title{Correcting the spectroscopic surface gravity using transits and asteroseismology}

   \subtitle{No significant effect on temperatures or metallicities with ARES+MOOG in LTE}

   \author{A. Mortier\inst{1,2},
           S.G. Sousa\inst{1,3,4},
           V.Zh. Adibekyan\inst{1},
           I.M. Brand\~ao\inst{1},
	\and
           N.C. Santos\inst{1,3}
          }

   \institute{Instituto de Astrof\'{\i}sica e Ci\^encias do Espa\c co, Universidade do Porto, CAUP, Rua das Estrelas, 4150-762 Porto, Portugal\\
              \email{am352@st-andrews.ac.uk}
	\and
	SUPA, School of Physics and Astronomy, University of St Andrews, St Andrews KY16 9SS, UK
	      \and
	      Departamento de F\'{\i}sica e Astronomia, Faculdade de Ci\^encias, Universidade do Porto, Portugal	      
	\and
	      Instituto de Astrof\'{\i}sica de Canarias, 38200 La Laguna, Tenerife, Spain 
	      }

   \date{Received 04 July 2014; Accepted 02 October 2014}

 
  \abstract
   {Precise stellar parameters (effective temperature, surface gravity, metallicity, stellar mass, and radius) are crucial for several reasons, amongst which are the precise characterization of orbiting exoplanets and the correct determination of galactic chemical evolution. The atmospheric parameters are extremely important because all the other stellar parameters depend on them. Using our standard equivalent-width method on high-resolution spectroscopy, good precision can be obtained for the derived effective temperature and metallicity. The surface gravity, however, is usually not well constrained with spectroscopy.}
   {We use two different samples of FGK dwarfs to study the effect of the stellar surface gravity on the precise spectroscopic determination of the other atmospheric parameters. Furthermore, we present a straightforward formula for correcting the spectroscopic surface gravities derived by our method and with our linelists.}
   {Our spectroscopic analysis is based on Kurucz models in LTE, performed with the MOOG code to derive the atmospheric parameters. The surface gravity was either left free or fixed to a predetermined value. The latter is either obtained through a photometric transit light curve or derived using asteroseismology.}
   {We find first that, despite some minor trends, the effective temperatures and metallicities for FGK dwarfs derived with the described method and linelists are, in most cases, only affected within the errorbars by using different values for the surface gravity, even for very large differences in surface gravity, so they can be trusted. The temperatures derived with a fixed surface gravity continue to be compatible within 1 sigma with the accurate results of the InfraRed Flux Method (IRFM), as is the case for the unconstrained temperatures. Secondly, we find that the spectroscopic surface gravity can easily be corrected to a more accurate value using a linear function with the effective temperature.}
   {}

   \authorrunning{Mortier, A. et al.}
   \maketitle
%

\section{Introduction}

Precise stellar parameters, such as effective temperature, surface gravity, metallicity, stellar mass, and stellar radius, are crucial for several reasons in astronomy. Amongst these, there are the precise characterization of planetary systems \citep[e.g.][]{Tor12,ME13c}, discovery of the possible link between the properties of stars and the existence of a planet \citep[e.g.][]{Adi13,Bea13,ME13}, and the complete and accurate picture of Galactic evolution \citep[e.g.][]{Edv93,McW08,Min13}.

In the ever-growing exoplanetary field\footnote{More than 1700 discovered exoplanets, see \url{www.exoplanet.eu}}, accurate and precise stellar parameters are necessary for the precise characterization of exoplanets. The main bulk of the discovered exoplanets has been found using radial velocities and/or the photometric transit technique. Separately, these techniques only partly characterize the planet. With radial velocities, a constraint is put on the planetary mass ($M_p\sin i$), while the transit technique is used to determine the planetary radius ($R_p$). Good knowledge of both these properties is essential for understanding the different kinds of planets and their distributions in the Galaxy \citep[e.g.][]{Buch14,Dum14,Marcy14}.

However, these planetary characteristics (mass, radius, and thus mean density) are highly dependent on the knowledge of the stellar characteristics ($M_p \propto M_{\ast}^{2/3}$ and $R_p \propto R_{\ast}$) \citep[e.g. ][]{Tor12,ME13c}. The stellar mass and radius, in turn, depend on the effective temperature, surface gravity, and the metallicity of the star, therefore it is extremely important to obtain precise atmospheric stellar properties.

Furthermore, to minimize the errors and to obtain comparable results, a uniform analysis is required \citep{Tor08,Tor12,San13} to guarantee the best possible homogeneity in the results. By homogeneously deriving precise stellar parameters we also gain more than just improving planetary parameters. Observational and theoretical works have shown that the processes of planet formation and evolution seem to depend on several stellar properties, such as stellar metallicity and mass \citep[e.g. ][]{But06,Udry07,Bow10,John10,Mayor11,Sou11b,Mor12,ME13,Adi13}. With large samples of planet hosts with homogeneously derived stellar and planetary parameters, we can look for correlations between the various parameters and statistically evaluate them. These correlations will allow us to narrow down the theories of planet formation.

Not just exoplanetary science benefits from having precise, accurate, and homogeneous stellar properties. These can also be useful to explain the formation and evolution of stars and thus of our Galaxy, which consists of different structures all with different properties. It has been shown for example that there is a difference in metallicity (iron and other heavy elements) between the thin disk and the thick disk \citep[e.g.][]{Edv93,Bens05,Hay08,Adi13b}. To properly understand the different stellar populations and their origins in the Milky Way, we need precise and homogeneous stellar parameters.

To derive a set of precise stellar properties (effective temperature $T_{\mathrm{eff}}$, surface gravity $\log g$, metallicity [Fe/H], and microturbulence $\xi$), high-resolution spectroscopy is usually the best approach. Commonly, two methods are used to analyse these spectra: spectral synthesis and spectral line analysis. The first method compares observed spectra with synthetic ones, for example with the code SME \citep{Val96} or MATISSE \citep{Rec06}. Spectral line analysis, as used in this work, makes use of the equivalent width (EW) of absorption lines (usually the \ion{Fe}{i} and \ion{Fe}{ii} lines) to demand excitation and ionization equilibrium.

\begin{figure*}[th!]
\begin{center}
\includegraphics[width=5.8cm]{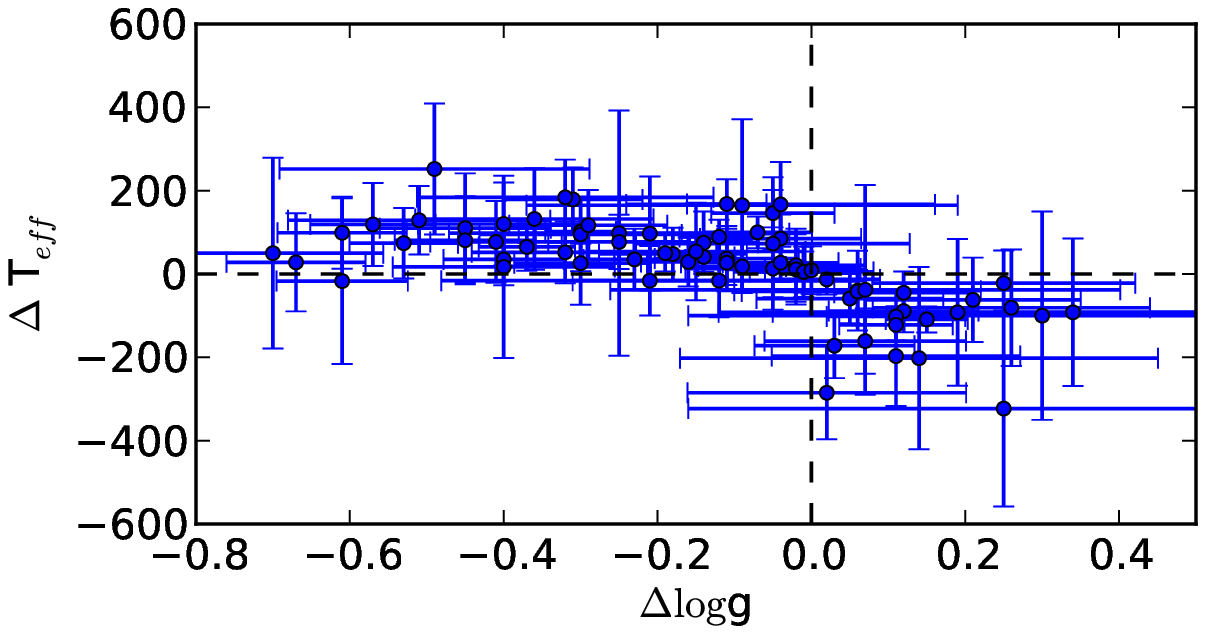}
\includegraphics[width=5.8cm]{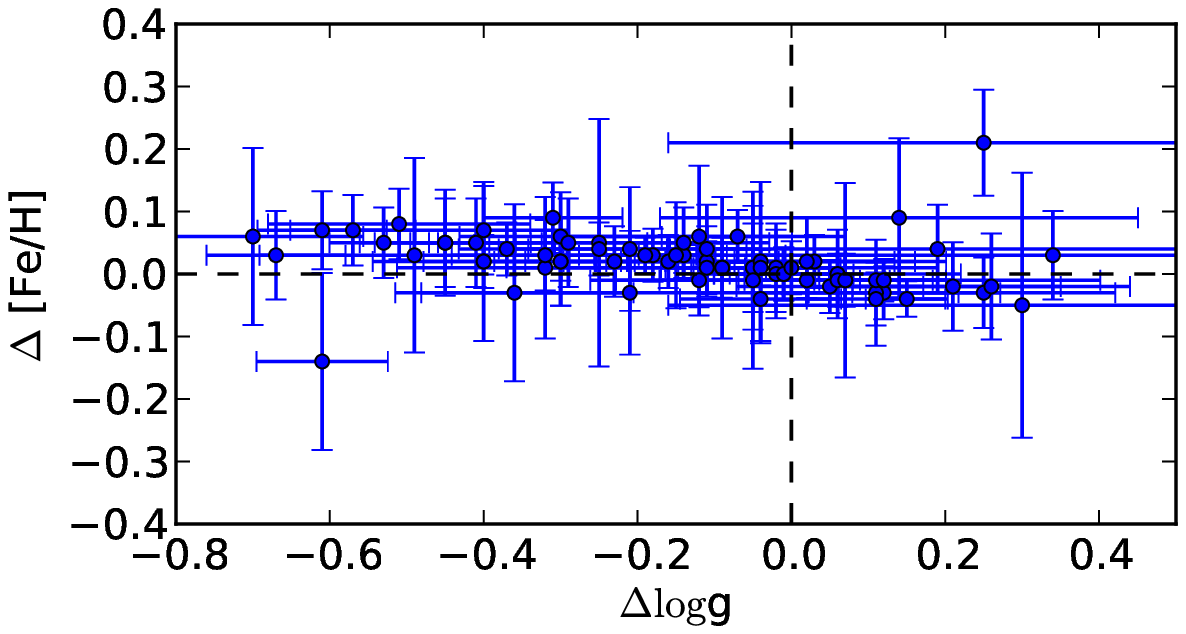}
\includegraphics[width=5.8cm]{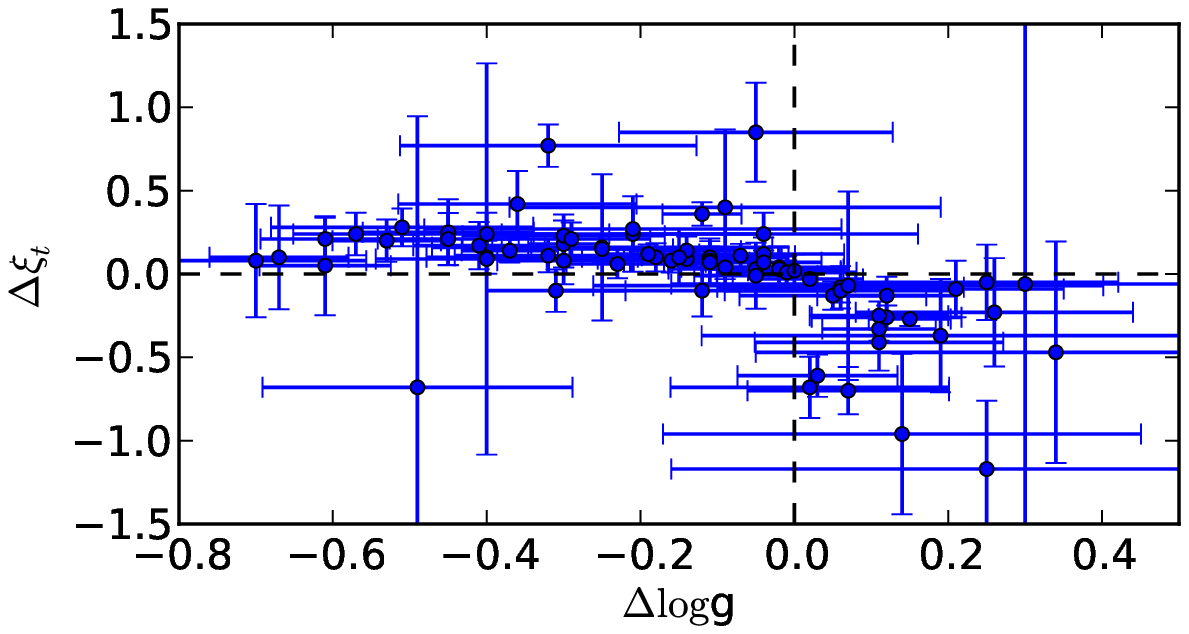}
\caption{Differences of the spectroscopic results (left to right: effective temperature, metallicity, and microturbulence) as a function of the difference in $\log$g (defined as `constrained with transit $\log g$ - unconstrained').}
\label{FigFL1}
\end{center}
\end{figure*}

Both methods have been shown to provide surface gravities that are not well constrained and do not compare well with surface gravities as obtained from other non-spectroscopic methods, such as asteroseismology or stellar models \citep[e.g. ][]{Tor12,Hub13,ME13c}. This surface gravity is important for the determination of the stellar mass and especially the stellar radius as shown in \citet{ME13c}.

In this work, we take a closer look at the surface gravity and its effect on the determination of the other atmospheric parameters. In Section \ref{Meth}, we present the uniform spectroscopic method we use. Section \ref{Tra} handles the effect of fixing the surface gravity to a value obtained by transit photometry and a possible correction formula. The same study is then done for the more accurate surface gravities as obtained by asteroseismolgy (Section \ref{Seis}). Finally, we discuss in Section \ref{Dis}.


\section{Spectroscopic method}\label{Meth}

Over the years, we have developed a homogeneous method to derive stellar parameters \citep[e.g.][]{San04,Sou08,Sou11b,Tsa13}. This method is based on the analysis of iron lines from high-resolution spectra. Details of this method can be found in \citet{San13} and references therein. Here we only give an overview of the method.

EWs of iron lines (\ion{Fe}{I} and \ion{Fe}{II}) are automatically calculated with the code ARES \citep[Automatic Routine for line Equivalent widths in stellar Spectra -][]{Sou07} for which the large lists with stable lines of \citet{Sou08} and \citet{Tsa13} are used for stars hotter and cooler than 5200\,K, respectively. These EWs are then used together with a grid of ATLAS plane-parallel model atmospheres \citep{Kur93} to determine the atmospheric stellar parameters, $T_{\mathrm{eff}}$, $\log g$, [Fe/H], and $\xi$. Therefore, we use the MOOG code\footnote{\url{http://www.as.utexas.edu/~chris/moog.html}} \citep{Sne73} in which we assume Local Thermodynamic Equilibrium (LTE).

By imposing excitation and ionization equilibrium, the atmospheric parameters are determined using an iterative minimization code based on the Downhill Simplex Method \citep{Pre92}. 

The same method can be used whilst fixing the surface gravity to a predetermined value (see next Sections). In this case however, ionization equilibrium will not be imposed as this is the main condition for determining the surface gravity. As a direct result, we do not use the \ion{Fe}{II} lines anymore. The value for the metallicity is thus determined by only using the \ion{Fe}{I} lines.

\begin{longtab}
\begin{longtable}{cccccccc}
\caption{\label{TabPar} Stellar (unconstrained) spectroscopic parameters used in this work. The last two columns contain the surface gravities as calculated with Equation \ref{EqFit}, resp. Equation \ref{EqFit2}}\\
\hline\hline
Name & T$_{eff,spec}$ & $\log g_{spec}$ & [Fe/H]$_{spec}$ & $\xi_{spec}$ & Ref. & $\log g_{corr,1}$ & $\log g_{corr,2}$ \\
 & (K) & (dex) & (dex) & (km s$^{-1}$) & & (dex) & (dex) \\
\hline
\endfirsthead
\caption{continued.}\\
\hline\hline
Name & T$_{eff,spec}$ & $\log g_{spec}$ & [Fe/H]$_{spec}$ & $\xi_{spec}$ & Ref. & $\log g_{corr,1}$ & $\log g_{corr,2}$ \\
 & (K) & (dex) & (dex) & (km s$^{-1}$) & & (dex) & (dex) \\
\hline
\endhead
\hline
\endfoot
\object{CoRoT-1} & 6397 $\pm$ 54 &   4.66 $\pm$   0.09 &   0.03 $\pm$   0.04 &   1.68 $\pm$   0.09 & (1) &   4.32 $\pm$   0.24 &   4.27 $\pm$   0.22 \\
\object{CoRoT-10} & 5025 $\pm$ 155 &   4.47 $\pm$   0.31 &   0.06 $\pm$   0.09 &   1.26 $\pm$   0.34 & (1) &   4.76 $\pm$   0.37 &   4.62 $\pm$   0.36 \\
\object{CoRoT-12} & 5715 $\pm$ 208 &   4.66 $\pm$   0.22 &   0.17 $\pm$   0.14 &   1.07 $\pm$   0.31 & (1) &   4.63 $\pm$   0.32 &   4.54 $\pm$   0.30 \\
\object{CoRoT-2} & 5697 $\pm$ 97 &   4.73 $\pm$   0.17 &  -0.09 $\pm$   0.07 &   1.64 $\pm$   0.16 & (1) &   4.71 $\pm$   0.27 &   4.61 $\pm$   0.26 \\
\object{CoRoT-4} & 6344 $\pm$ 93 &   4.82 $\pm$   0.11 &   0.15 $\pm$   0.06 &   1.74 $\pm$   0.14 & (1) &   4.50 $\pm$   0.25 &   4.45 $\pm$   0.23 \\
\object{CoRoT-5} & 6240 $\pm$ 70 &   4.46 $\pm$   0.11 &   0.04 $\pm$   0.05 &   1.28 $\pm$   0.09 & (1) &   4.19 $\pm$   0.25 &   4.13 $\pm$   0.23 \\
\object{CoRoT-7} & 5288 $\pm$ 27 &   4.40 $\pm$   0.07 &   0.02 $\pm$   0.02 &   0.90 $\pm$   0.05 & (1) &   4.57 $\pm$   0.21 &   4.44 $\pm$   0.20 \\
\object{CoRoT-8} & 5143 $\pm$ 178 &   4.42 $\pm$   0.33 &   0.22 $\pm$   0.11 &   0.61 $\pm$   0.40 & (1) &   4.65 $\pm$   0.39 &   4.52 $\pm$   0.38 \\
\object{CoRoT-9} & 5613 $\pm$ 36 &   4.35 $\pm$   0.09 &  -0.02 $\pm$   0.03 &   0.90 $\pm$   0.05 & (1) &   4.37 $\pm$   0.23 &   4.27 $\pm$   0.21 \\
\object{HAT-P-1} & 6076 $\pm$ 27 &   4.47 $\pm$   0.07 &   0.21 $\pm$   0.03 &   1.17 $\pm$   0.05 & (1) &   4.28 $\pm$   0.23 &   4.21 $\pm$   0.21 \\
\object{HAT-P-11} & 4624 $\pm$ 225 &   4.15 $\pm$   0.59 &   0.26 $\pm$   0.08 &   0.39 $\pm$   0.70 & (1) &   4.62 $\pm$   0.63 &   4.45 $\pm$   0.62 \\
\object{HAT-P-17} & 5332 $\pm$ 55 &   4.45 $\pm$   0.13 &   0.05 $\pm$   0.03 &   0.82 $\pm$   0.10 & (1) &   4.60 $\pm$   0.24 &   4.48 $\pm$   0.23 \\
\object{HAT-P-20} & 4502 $\pm$ 188 &   4.32 $\pm$   0.60 &   0.12 $\pm$   0.15 &   0.73 $\pm$   0.60 & (1) &   4.85 $\pm$   0.64 &   4.67 $\pm$   0.63 \\
\object{HAT-P-26} & 5011 $\pm$ 55 &   4.31 $\pm$   0.17 &   0.01 $\pm$   0.04 &   0.48 $\pm$   0.16 & (1) &   4.60 $\pm$   0.26 &   4.46 $\pm$   0.25 \\
\object{HAT-P-27} & 5316 $\pm$ 55 &   4.48 $\pm$   0.10 &   0.30 $\pm$   0.03 &   0.82 $\pm$   0.09 & (1) &   4.63 $\pm$   0.23 &   4.51 $\pm$   0.21 \\
\object{HAT-P-30} & 6338 $\pm$ 42 &   4.52 $\pm$   0.06 &   0.12 $\pm$   0.03 &   1.40 $\pm$   0.05 & (1) &   4.21 $\pm$   0.23 &   4.16 $\pm$   0.21 \\
\object{HAT-P-35} & 6178 $\pm$ 45 &   4.40 $\pm$   0.09 &   0.12 $\pm$   0.03 &   1.34 $\pm$   0.06 & (1) &   4.16 $\pm$   0.24 &   4.10 $\pm$   0.22 \\
\object{HAT-P-4} & 6054 $\pm$ 60 &   4.17 $\pm$   0.28 &   0.35 $\pm$   0.08 &   1.59 $\pm$   0.09 & (1) &   3.99 $\pm$   0.35 &   3.92 $\pm$   0.34 \\
\object{HAT-P-6} & 6855 $\pm$ 111 &   4.69 $\pm$   0.20 &  -0.08 $\pm$   0.11 &   2.85 $\pm$   1.15 & (1) &   4.14 $\pm$   0.31 &   4.12 $\pm$   0.29 \\
\object{HAT-P-7} & 6525 $\pm$ 61 &   4.09 $\pm$   0.08 &   0.31 $\pm$   0.07 &   1.78 $\pm$   0.14 & (1) &   3.69 $\pm$   0.24 &   3.65 $\pm$   0.22 \\
\object{HAT-P-8} & 6550 $\pm$ 61 &   4.80 $\pm$   0.08 &   0.07 $\pm$   0.04 &   1.93 $\pm$   0.09 & (1) &   4.39 $\pm$   0.24 &   4.35 $\pm$   0.22 \\
\object{HD149026} & 6162 $\pm$ 41 &   4.37 $\pm$   0.10 &   0.36 $\pm$   0.05 &   1.41 $\pm$   0.07 & (1) &   4.14 $\pm$   0.24 &   4.07 $\pm$   0.22 \\
\object{HD17156} & 6084 $\pm$ 29 &   4.33 $\pm$   0.05 &   0.23 $\pm$   0.04 &   1.47 $\pm$   0.05 & (1) &   4.13 $\pm$   0.22 &   4.06 $\pm$   0.21 \\
\object{HD189733} & 5109 $\pm$ 146 &   4.69 $\pm$   0.28 &   0.03 $\pm$   0.08 &   0.78 $\pm$   0.33 & (1) &   4.94 $\pm$   0.35 &   4.80 $\pm$   0.34 \\
\object{HD209458} & 6118 $\pm$ 25 &   4.50 $\pm$   0.04 &   0.03 $\pm$   0.02 &   1.21 $\pm$   0.03 & (1) &   4.29 $\pm$   0.22 &   4.22 $\pm$   0.20 \\
\object{HD80606} & 5574 $\pm$ 72 &   4.46 $\pm$   0.20 &   0.32 $\pm$   0.09 &   1.14 $\pm$   0.09 & (1) &   4.50 $\pm$   0.29 &   4.39 $\pm$   0.28 \\
\object{HD97658} & 5137 $\pm$ 36 &   4.47 $\pm$   0.09 &  -0.35 $\pm$   0.02 &   0.63 $\pm$   0.08 & (1) &   4.71 $\pm$   0.22 &   4.57 $\pm$   0.21 \\
\object{Kepler-17} & 5781 $\pm$ 85 &   4.53 $\pm$   0.12 &   0.26 $\pm$   0.10 &   1.73 $\pm$   0.14 & (1) &   4.47 $\pm$   0.25 &   4.38 $\pm$   0.23 \\
\object{Kepler-21} & 6409 $\pm$ 44 &   4.43 $\pm$   0.06 &  -0.03 $\pm$   0.03 &   1.86 $\pm$   0.07 & (1) &   4.08 $\pm$   0.23 &   4.04 $\pm$   0.21 \\
\object{KOI-135} & 6041 $\pm$ 143 &   4.26 $\pm$   0.05 &   0.33 $\pm$   0.11 &   1.85 $\pm$   0.26 & (1) &   4.08 $\pm$   0.23 &   4.01 $\pm$   0.21 \\
\object{KOI-204} & 5757 $\pm$ 134 &   4.15 $\pm$   0.06 &   0.26 $\pm$   0.10 &   1.75 $\pm$   0.19 & (1) &   4.10 $\pm$   0.23 &   4.01 $\pm$   0.21 \\
\object{OGLE-TR-10} & 6075 $\pm$ 86 &   4.54 $\pm$   0.15 &   0.28 $\pm$   0.10 &   1.45 $\pm$   0.14 & (1) &   4.35 $\pm$   0.27 &   4.28 $\pm$   0.25 \\
\object{OGLE-TR-111} & 4800 $\pm$ 177 &   4.24 $\pm$   0.46 &   0.22 $\pm$   0.15 &   0.30 $\pm$   1.38 & (1) &   4.63 $\pm$   0.51 &   4.47 $\pm$   0.50 \\
\object{OGLE-TR-113} & 4781 $\pm$ 166 &   4.31 $\pm$   0.41 &   0.03 $\pm$   0.06 &   1.24 $\pm$   0.29 & (1) &   4.71 $\pm$   0.46 &   4.55 $\pm$   0.45 \\
\object{OGLE-TR-132} & 6210 $\pm$ 59 &   4.51 $\pm$   0.27 &   0.37 $\pm$   0.07 &   1.23 $\pm$   0.09 & (1) &   4.26 $\pm$   0.35 &   4.20 $\pm$   0.34 \\
\object{OGLE-TR-182} & 5924 $\pm$ 64 &   4.47 $\pm$   0.18 &   0.37 $\pm$   0.08 &   0.91 $\pm$   0.09 & (1) &   4.35 $\pm$   0.28 &   4.27 $\pm$   0.27 \\
\object{OGLE-TR-211} & 6325 $\pm$ 91 &   4.22 $\pm$   0.17 &   0.11 $\pm$   0.10 &   1.63 $\pm$   0.21 & (1) &   3.91 $\pm$   0.28 &   3.86 $\pm$   0.27 \\
\object{OGLE-TR-56} & 6119 $\pm$ 62 &   4.21 $\pm$   0.19 &   0.25 $\pm$   0.08 &   1.48 $\pm$   0.11 & (1) &   4.00 $\pm$   0.29 &   3.93 $\pm$   0.28 \\
\object{TrES-1} & 5226 $\pm$ 38 &   4.40 $\pm$   0.10 &   0.06 $\pm$   0.05 &   0.90 $\pm$   0.05 & (1) &   4.60 $\pm$   0.22 &   4.47 $\pm$   0.21 \\
\object{TrES-2} & 5795 $\pm$ 73 &   4.30 $\pm$   0.13 &   0.06 $\pm$   0.08 &   0.79 $\pm$   0.12 & (1) &   4.24 $\pm$   0.25 &   4.15 $\pm$   0.24 \\
\object{TrES-3} & 5502 $\pm$ 157 &   4.44 $\pm$   0.22 &  -0.10 $\pm$   0.19 &   1.00 $\pm$   0.30 & (1) &   4.51 $\pm$   0.31 &   4.40 $\pm$   0.30 \\
\object{TrES-4} & 6293 $\pm$ 96 &   4.20 $\pm$   0.27 &   0.34 $\pm$   0.10 &   2.01 $\pm$   0.17 & (1) &   3.91 $\pm$   0.35 &   3.85 $\pm$   0.34 \\
\object{WASP-1} & 6252 $\pm$ 45 &   4.32 $\pm$   0.05 &   0.23 $\pm$   0.03 &   1.42 $\pm$   0.05 & (1) &   4.05 $\pm$   0.22 &   3.99 $\pm$   0.21 \\
\object{WASP-10} & 4645 $\pm$ 125 &   4.27 $\pm$   0.39 &   0.04 $\pm$   0.05 &   0.58 $\pm$   0.47 & (1) &   4.73 $\pm$   0.44 &   4.56 $\pm$   0.43 \\
\object{WASP-11} & 4881 $\pm$ 125 &   4.44 $\pm$   0.31 &   0.01 $\pm$   0.05 &   0.64 $\pm$   0.24 & (1) &   4.79 $\pm$   0.37 &   4.64 $\pm$   0.36 \\
\object{WASP-12} & 6313 $\pm$ 52 &   4.37 $\pm$   0.12 &   0.21 $\pm$   0.04 &   1.65 $\pm$   0.07 & (1) &   4.07 $\pm$   0.25 &   4.02 $\pm$   0.24 \\
\object{WASP-13} & 6025 $\pm$ 21 &   4.19 $\pm$   0.03 &   0.11 $\pm$   0.05 &   1.28 $\pm$   0.10 & (1) &   4.02 $\pm$   0.22 &   3.95 $\pm$   0.20 \\
\object{WASP-15} & 6573 $\pm$ 70 &   4.79 $\pm$   0.08 &   0.09 $\pm$   0.04 &   1.72 $\pm$   0.09 & (1) &   4.37 $\pm$   0.24 &   4.33 $\pm$   0.22 \\
\object{WASP-16} & 5726 $\pm$ 22 &   4.34 $\pm$   0.05 &   0.13 $\pm$   0.02 &   0.97 $\pm$   0.03 & (1) &   4.31 $\pm$   0.22 &   4.21 $\pm$   0.20 \\
\object{WASP-17} & 6794 $\pm$ 83 &   4.83 $\pm$   0.09 &  -0.12 $\pm$   0.05 &   2.57 $\pm$   0.22 & (1) &   4.31 $\pm$   0.25 &   4.29 $\pm$   0.23 \\
\object{WASP-18} & 6526 $\pm$ 69 &   4.73 $\pm$   0.08 &   0.19 $\pm$   0.05 &   1.83 $\pm$   0.10 & (1) &   4.33 $\pm$   0.24 &   4.29 $\pm$   0.22 \\
\object{WASP-19} & 5591 $\pm$ 62 &   4.46 $\pm$   0.09 &   0.26 $\pm$   0.05 &   1.23 $\pm$   0.09 & (1) &   4.49 $\pm$   0.23 &   4.39 $\pm$   0.21 \\
\object{WASP-2} & 5109 $\pm$ 72 &   4.33 $\pm$   0.14 &   0.02 $\pm$   0.05 &   0.57 $\pm$   0.12 & (1) &   4.58 $\pm$   0.25 &   4.44 $\pm$   0.23 \\
\object{WASP-21} & 5924 $\pm$ 55 &   4.39 $\pm$   0.09 &  -0.22 $\pm$   0.04 &   1.06 $\pm$   0.08 & (1) &   4.27 $\pm$   0.23 &   4.19 $\pm$   0.22 \\
\object{WASP-22} & 6153 $\pm$ 46 &   4.57 $\pm$   0.09 &   0.26 $\pm$   0.03 &   1.36 $\pm$   0.06 & (1) &   4.34 $\pm$   0.24 &   4.28 $\pm$   0.22 \\
\object{WASP-23} & 5046 $\pm$ 99 &   4.33 $\pm$   0.18 &   0.05 $\pm$   0.06 &   0.64 $\pm$   0.23 & (1) &   4.61 $\pm$   0.27 &   4.47 $\pm$   0.26 \\
\object{WASP-24} & 6297 $\pm$ 58 &   4.76 $\pm$   0.17 &   0.09 $\pm$   0.04 &   1.41 $\pm$   0.08 & (1) &   4.47 $\pm$   0.28 &   4.41 $\pm$   0.27 \\
\object{WASP-25} & 5736 $\pm$ 35 &   4.52 $\pm$   0.09 &   0.06 $\pm$   0.03 &   1.11 $\pm$   0.05 & (1) &   4.48 $\pm$   0.23 &   4.39 $\pm$   0.21 \\
\object{WASP-26} & 6034 $\pm$ 31 &   4.44 $\pm$   0.06 &   0.16 $\pm$   0.02 &   1.28 $\pm$   0.04 & (1) &   4.27 $\pm$   0.22 &   4.19 $\pm$   0.21 \\
\object{WASP-28} & 6134 $\pm$ 38 &   4.55 $\pm$   0.05 &  -0.12 $\pm$   0.03 &   1.17 $\pm$   0.06 & (1) &   4.33 $\pm$   0.22 &   4.26 $\pm$   0.21 \\
\object{WASP-29} & 5203 $\pm$ 102 &   4.93 $\pm$   0.21 &   0.17 $\pm$   0.05 &   1.77 $\pm$   0.22 & (1) &   5.14 $\pm$   0.29 &   5.01 $\pm$   0.28 \\
\object{WASP-31} & 6443 $\pm$ 75 &   4.76 $\pm$   0.09 &  -0.08 $\pm$   0.05 &   1.62 $\pm$   0.11 & (1) &   4.40 $\pm$   0.24 &   4.35 $\pm$   0.23 \\
\object{WASP-32} & 6427 $\pm$ 141 &   4.93 $\pm$   0.08 &   0.28 $\pm$   0.10 &   1.20 $\pm$   0.21 & (1) &   4.58 $\pm$   0.24 &   4.53 $\pm$   0.23 \\
\object{WASP-34} & 5704 $\pm$ 26 &   4.35 $\pm$   0.05 &   0.08 $\pm$   0.02 &   0.97 $\pm$   0.03 & (1) &   4.33 $\pm$   0.21 &   4.23 $\pm$   0.20 \\
\object{WASP-35} & 6072 $\pm$ 62 &   4.69 $\pm$   0.13 &  -0.05 $\pm$   0.05 &   1.26 $\pm$   0.09 & (1) &   4.50 $\pm$   0.25 &   4.43 $\pm$   0.24 \\
\object{WASP-36} & 5928 $\pm$ 59 &   4.51 $\pm$   0.09 &  -0.01 $\pm$   0.05 &   0.89 $\pm$   0.09 & (1) &   4.38 $\pm$   0.23 &   4.30 $\pm$   0.22 \\
\object{WASP-38} & 6436 $\pm$ 60 &   4.80 $\pm$   0.07 &   0.06 $\pm$   0.04 &   1.75 $\pm$   0.09 & (1) &   4.44 $\pm$   0.23 &   4.40 $\pm$   0.22 \\
\object{WASP-4} & 5513 $\pm$ 43 &   4.50 $\pm$   0.07 &   0.03 $\pm$   0.03 &   0.86 $\pm$   0.07 & (1) &   4.56 $\pm$   0.22 &   4.46 $\pm$   0.20 \\
\object{WASP-41} & 5546 $\pm$ 33 &   4.53 $\pm$   0.07 &   0.06 $\pm$   0.02 &   1.08 $\pm$   0.05 & (1) &   4.58 $\pm$   0.22 &   4.47 $\pm$   0.20 \\
\object{WASP-42} & 5315 $\pm$ 79 &   4.50 $\pm$   0.18 &   0.29 $\pm$   0.05 &   1.16 $\pm$   0.13 & (1) &   4.65 $\pm$   0.27 &   4.53 $\pm$   0.26 \\
\object{WASP-47} & 5576 $\pm$ 68 &   4.28 $\pm$   0.16 &   0.36 $\pm$   0.05 &   1.25 $\pm$   0.09 & (1) &   4.32 $\pm$   0.26 &   4.21 $\pm$   0.25 \\
\object{WASP-5} & 5785 $\pm$ 83 &   4.54 $\pm$   0.14 &   0.17 $\pm$   0.06 &   0.96 $\pm$   0.12 & (1) &   4.48 $\pm$   0.26 &   4.39 $\pm$   0.24 \\
\object{WASP-50} & 5518 $\pm$ 42 &   4.43 $\pm$   0.12 &   0.13 $\pm$   0.03 &   1.25 $\pm$   0.06 & (1) &   4.49 $\pm$   0.24 &   4.38 $\pm$   0.23 \\
\object{WASP-54} & 6296 $\pm$ 40 &   4.37 $\pm$   0.06 &   0.00 $\pm$   0.03 &   1.45 $\pm$   0.05 & (1) &   4.08 $\pm$   0.23 &   4.02 $\pm$   0.21 \\
\object{WASP-55} & 6070 $\pm$ 53 &   4.55 $\pm$   0.07 &   0.09 $\pm$   0.04 &   1.10 $\pm$   0.06 & (1) &   4.36 $\pm$   0.23 &   4.29 $\pm$   0.21 \\
\object{WASP-6} & 5383 $\pm$ 41 &   4.52 $\pm$   0.06 &  -0.14 $\pm$   0.03 &   0.80 $\pm$   0.07 & (1) &   4.64 $\pm$   0.21 &   4.53 $\pm$   0.20 \\
\object{WASP-62} & 6391 $\pm$ 70 &   4.73 $\pm$   0.11 &   0.24 $\pm$   0.05 &   1.50 $\pm$   0.09 & (1) &   4.39 $\pm$   0.25 &   4.34 $\pm$   0.23 \\
\object{WASP-63} & 5715 $\pm$ 60 &   4.29 $\pm$   0.10 &   0.28 $\pm$   0.05 &   1.28 $\pm$   0.07 & (1) &   4.26 $\pm$   0.23 &   4.17 $\pm$   0.22 \\
\object{WASP-66} & 7051 $\pm$ 79 &   5.00 $\pm$   0.08 &   0.05 $\pm$   0.05 &   3.07 $\pm$   0.27 & (1) &   4.36 $\pm$   0.25 &   4.36 $\pm$   0.23 \\
\object{WASP-67} & 5417 $\pm$ 85 &   4.40 $\pm$   0.16 &   0.18 $\pm$   0.06 &   1.16 $\pm$   0.12 & (1) &   4.51 $\pm$   0.26 &   4.39 $\pm$   0.25 \\
\object{WASP-7} & 6621 $\pm$ 155 &   4.62 $\pm$   0.14 &   0.12 $\pm$   0.09 &   3.00 $\pm$   0.83 & (1) &   4.18 $\pm$   0.27 &   4.15 $\pm$   0.26 \\
\object{WASP-71} & 6180 $\pm$ 52 &   4.15 $\pm$   0.06 &   0.37 $\pm$   0.04 &   1.69 $\pm$   0.06 & (1) &   3.91 $\pm$   0.23 &   3.85 $\pm$   0.21 \\
\object{WASP-77A} & 5605 $\pm$ 41 &   4.37 $\pm$   0.09 &   0.07 $\pm$   0.03 &   1.09 $\pm$   0.06 & (1) &   4.39 $\pm$   0.23 &   4.29 $\pm$   0.21 \\
\object{WASP-78} & 6291 $\pm$ 71 &   4.19 $\pm$   0.08 &  -0.07 $\pm$   0.05 &   1.63 $\pm$   0.10 & (1) &   3.90 $\pm$   0.24 &   3.84 $\pm$   0.22 \\
\object{WASP-79} & 7002 $\pm$ 162 &   4.77 $\pm$   0.14 &   0.19 $\pm$   0.10 &   2.64 $\pm$   0.24 & (1) &   4.15 $\pm$   0.28 &   4.15 $\pm$   0.26 \\
\object{WASP-8} & 5690 $\pm$ 36 &   4.42 $\pm$   0.15 &   0.29 $\pm$   0.03 &   1.25 $\pm$   0.05 & (1) &   4.40 $\pm$   0.26 &   4.31 $\pm$   0.24 \\
\object{XO-1} & 5754 $\pm$ 42 &   4.61 $\pm$   0.05 &  -0.01 $\pm$   0.05 &   1.07 $\pm$   0.09 & (1) &   4.56 $\pm$   0.22 &   4.47 $\pm$   0.20 \\
\object{KIC1430163} & 6833 $\pm$ 87 &   4.70 $\pm$   0.11 &   0.02 $\pm$   0.06 &   2.12 $\pm$   0.10 & (2) &   4.16 $\pm$   0.26 &   4.14 $\pm$   0.24 \\
\object{KIC1435467} & 6485 $\pm$ 92 &   4.53 $\pm$   0.13 &   0.08 $\pm$   0.07 &   2.02 $\pm$   0.09 & (2) &   4.15 $\pm$   0.26 &   4.11 $\pm$   0.25 \\
\object{KIC3427720} & 6111 $\pm$ 68 &   4.51 $\pm$   0.11 &   0.04 $\pm$   0.06 &   1.25 $\pm$   0.04 & (2) &   4.30 $\pm$   0.24 &   4.23 $\pm$   0.23 \\
\object{KIC3456181} & 6584 $\pm$ 91 &   4.43 $\pm$   0.11 &  -0.02 $\pm$   0.07 &   2.01 $\pm$   0.11 & (2) &   4.01 $\pm$   0.25 &   3.97 $\pm$   0.24 \\
\object{KIC3632418} & 6409 $\pm$ 74 &   4.43 $\pm$   0.12 &  -0.03 $\pm$   0.06 &   1.86 $\pm$   0.06 & (2) &   4.09 $\pm$   0.25 &   4.04 $\pm$   0.24 \\
\object{KIC3643774} & 6125 $\pm$ 75 &   4.39 $\pm$   0.12 &   0.25 $\pm$   0.06 &   1.39 $\pm$   0.05 & (2) &   4.18 $\pm$   0.25 &   4.11 $\pm$   0.23 \\
\object{KIC3656476} & 5719 $\pm$ 64 &   4.26 $\pm$   0.11 &   0.28 $\pm$   0.05 &   1.11 $\pm$   0.03 & (2) &   4.23 $\pm$   0.24 &   4.14 $\pm$   0.22 \\
\object{KIC4072740} & 4960 $\pm$ 77 &   3.49 $\pm$   0.13 &   0.19 $\pm$   0.06 &   1.13 $\pm$   0.06 & (2) &   3.81 $\pm$   0.24 &   3.66 $\pm$   0.23 \\
\object{KIC4346201} & 6239 $\pm$ 91 &   4.28 $\pm$   0.12 &  -0.17 $\pm$   0.07 &   1.64 $\pm$   0.10 & (2) &   4.01 $\pm$   0.25 &   3.95 $\pm$   0.24 \\
\object{KIC4586099} & 6533 $\pm$ 80 &   4.37 $\pm$   0.11 &  -0.04 $\pm$   0.06 &   1.84 $\pm$   0.08 & (2) &   3.97 $\pm$   0.25 &   3.93 $\pm$   0.24 \\
\object{KIC4638884} & 6684 $\pm$ 98 &   4.58 $\pm$   0.17 &  -0.05 $\pm$   0.08 &   3.39 $\pm$   0.28 & (2) &   4.11 $\pm$   0.29 &   4.08 $\pm$   0.27 \\
\object{KIC4914923} & 5948 $\pm$ 65 &   4.34 $\pm$   0.12 &   0.18 $\pm$   0.05 &   1.26 $\pm$   0.03 & (2) &   4.21 $\pm$   0.25 &   4.13 $\pm$   0.23 \\
\object{KIC4931390} & 6862 $\pm$ 80 &   4.55 $\pm$   0.11 &  -0.02 $\pm$   0.06 &   1.93 $\pm$   0.09 & (2) &   4.00 $\pm$   0.26 &   3.98 $\pm$   0.24 \\
\object{KIC5184732}\_esp & 5894 $\pm$ 68 &   4.31 $\pm$   0.12 &   0.43 $\pm$   0.06 &   1.18 $\pm$   0.03 & (2) &   4.20 $\pm$   0.25 &   4.12 $\pm$   0.23 \\
\object{KIC5184732}\_nar & 5877 $\pm$ 68 &   4.34 $\pm$   0.11 &   0.40 $\pm$   0.06 &   1.14 $\pm$   0.03 & (2) &   4.24 $\pm$   0.24 &   4.15 $\pm$   0.23 \\
\object{KIC5371516} & 6526 $\pm$ 107 &   4.49 $\pm$   0.15 &   0.11 $\pm$   0.08 &   2.35 $\pm$   0.14 & (2) &   4.09 $\pm$   0.27 &   4.05 $\pm$   0.26 \\
\object{KIC5450445} & 6396 $\pm$ 75 &   4.49 $\pm$   0.11 &   0.23 $\pm$   0.06 &   1.75 $\pm$   0.06 & (2) &   4.15 $\pm$   0.25 &   4.10 $\pm$   0.23 \\
\object{KIC5512589} & 5812 $\pm$ 66 &   4.05 $\pm$   0.11 &   0.12 $\pm$   0.06 &   1.20 $\pm$   0.03 & (2) &   3.98 $\pm$   0.24 &   3.89 $\pm$   0.23 \\
\object{KIC5773345} & 6399 $\pm$ 71 &   4.36 $\pm$   0.11 &   0.30 $\pm$   0.06 &   1.92 $\pm$   0.05 & (2) &   4.02 $\pm$   0.25 &   3.97 $\pm$   0.23 \\
\object{KIC5955122} & 6092 $\pm$ 69 &   4.26 $\pm$   0.12 &  -0.06 $\pm$   0.06 &   1.66 $\pm$   0.05 & (2) &   4.06 $\pm$   0.25 &   3.99 $\pm$   0.23 \\
\object{KIC6116048} & 6152 $\pm$ 66 &   4.53 $\pm$   0.10 &  -0.14 $\pm$   0.05 &   1.36 $\pm$   0.04 & (2) &   4.30 $\pm$   0.24 &   4.24 $\pm$   0.23 \\
\object{KIC6225718} & 6366 $\pm$ 70 &   4.61 $\pm$   0.11 &  -0.07 $\pm$   0.06 &   1.50 $\pm$   0.05 & (2) &   4.29 $\pm$   0.25 &   4.23 $\pm$   0.23 \\
\object{KIC6442183} & 5738 $\pm$ 62 &   4.14 $\pm$   0.10 &  -0.12 $\pm$   0.05 &   1.15 $\pm$   0.02 & (2) &   4.10 $\pm$   0.23 &   4.01 $\pm$   0.22 \\
\object{KIC6603624} & 5718 $\pm$ 78 &   4.44 $\pm$   0.13 &   0.28 $\pm$   0.06 &   1.16 $\pm$   0.06 & (2) &   4.41 $\pm$   0.25 &   4.32 $\pm$   0.23 \\
\object{KIC6933899} & 5921 $\pm$ 65 &   4.12 $\pm$   0.11 &   0.04 $\pm$   0.06 &   1.29 $\pm$   0.03 & (2) &   4.00 $\pm$   0.24 &   3.92 $\pm$   0.23 \\
\object{KIC7103006} & 6685 $\pm$ 86 &   4.50 $\pm$   0.11 &   0.19 $\pm$   0.06 &   1.98 $\pm$   0.08 & (2) &   4.03 $\pm$   0.25 &   4.00 $\pm$   0.24 \\
\object{KIC7668623} & 6580 $\pm$ 112 &   4.56 $\pm$   0.15 &   0.03 $\pm$   0.08 &   2.54 $\pm$   0.21 & (2) &   4.14 $\pm$   0.27 &   4.10 $\pm$   0.26 \\
\object{KIC7680114} & 5955 $\pm$ 68 &   4.41 $\pm$   0.11 &   0.12 $\pm$   0.06 &   1.30 $\pm$   0.04 & (2) &   4.27 $\pm$   0.24 &   4.19 $\pm$   0.23 \\
\object{KIC7747078} & 6114 $\pm$ 78 &   4.37 $\pm$   0.12 &  -0.11 $\pm$   0.06 &   1.65 $\pm$   0.07 & (2) &   4.16 $\pm$   0.25 &   4.09 $\pm$   0.23 \\
\object{KIC7799349} & 5175 $\pm$ 84 &   3.81 $\pm$   0.15 &   0.24 $\pm$   0.07 &   1.31 $\pm$   0.07 & (2) &   4.03 $\pm$   0.25 &   3.90 $\pm$   0.24 \\
\object{KIC7940546}\_esp & 6427 $\pm$ 82 &   4.52 $\pm$   0.12 &  -0.11 $\pm$   0.06 &   2.09 $\pm$   0.09 & (2) &   4.17 $\pm$   0.25 &   4.12 $\pm$   0.24 \\
\object{KIC7940546}\_nar & 6472 $\pm$ 84 &   4.59 $\pm$   0.12 &  -0.11 $\pm$   0.06 &   2.32 $\pm$   0.12 & (2) &   4.22 $\pm$   0.26 &   4.17 $\pm$   0.24 \\
\object{KIC7976303} & 6203 $\pm$ 76 &   4.15 $\pm$   0.11 &  -0.41 $\pm$   0.06 &   1.62 $\pm$   0.07 & (2) &   3.90 $\pm$   0.25 &   3.84 $\pm$   0.23 \\
\object{KIC8006161}\_esp & 5431 $\pm$ 82 &   4.45 $\pm$   0.13 &   0.30 $\pm$   0.06 &   0.95 $\pm$   0.10 & (2) &   4.55 $\pm$   0.24 &   4.44 $\pm$   0.23 \\
\object{KIC8006161}\_nar & 5468 $\pm$ 77 &   4.41 $\pm$   0.13 &   0.29 $\pm$   0.06 &   1.07 $\pm$   0.07 & (2) &   4.50 $\pm$   0.24 &   4.38 $\pm$   0.23 \\
\object{KIC8026226} & 6469 $\pm$ 78 &   4.32 $\pm$   0.13 &  -0.13 $\pm$   0.06 &   2.72 $\pm$   0.18 & (2) &   3.95 $\pm$   0.26 &   3.90 $\pm$   0.25 \\
\object{KIC8179536} & 6536 $\pm$ 74 &   4.64 $\pm$   0.11 &   0.13 $\pm$   0.06 &   1.61 $\pm$   0.05 & (2) &   4.24 $\pm$   0.25 &   4.20 $\pm$   0.24 \\
\object{KIC8228742} & 6295 $\pm$ 76 &   4.42 $\pm$   0.11 &   0.00 $\pm$   0.06 &   1.71 $\pm$   0.06 & (2) &   4.13 $\pm$   0.25 &   4.07 $\pm$   0.23 \\
\object{KIC8379927}\_esp & 6225 $\pm$ 95 &   4.76 $\pm$   0.13 &  -0.23 $\pm$   0.07 &   2.01 $\pm$   0.13 & (2) &   4.50 $\pm$   0.26 &   4.44 $\pm$   0.24 \\
\object{KIC8379927}\_nar & 6202 $\pm$ 73 &   4.47 $\pm$   0.12 &  -0.20 $\pm$   0.06 &   0.95 $\pm$   0.05 & (2) &   4.22 $\pm$   0.25 &   4.16 $\pm$   0.24 \\
\object{KIC8394589} & 6231 $\pm$ 75 &   4.54 $\pm$   0.11 &  -0.24 $\pm$   0.06 &   1.36 $\pm$   0.07 & (2) &   4.28 $\pm$   0.25 &   4.22 $\pm$   0.23 \\
\object{KIC8524425} & 5664 $\pm$ 65 &   4.09 $\pm$   0.11 &   0.13 $\pm$   0.05 &   1.16 $\pm$   0.03 & (2) &   4.09 $\pm$   0.24 &   3.99 $\pm$   0.22 \\
\object{KIC8561221} & 5352 $\pm$ 68 &   3.80 $\pm$   0.11 &  -0.04 $\pm$   0.06 &   1.14 $\pm$   0.04 & (2) &   3.94 $\pm$   0.23 &   3.82 $\pm$   0.22 \\
\object{KIC8694723}\_nar & 6445 $\pm$ 80 &   4.55 $\pm$   0.11 &  -0.39 $\pm$   0.06 &   1.91 $\pm$   0.11 & (2) &   4.19 $\pm$   0.25 &   4.14 $\pm$   0.23 \\
\object{KIC8694723}\_fies & 6489 $\pm$ 85 &   4.50 $\pm$   0.13 &  -0.35 $\pm$   0.06 &   1.98 $\pm$   0.13 & (2) &   4.12 $\pm$   0.26 &   4.08 $\pm$   0.25 \\
\object{KIC8702606} & 5578 $\pm$ 62 &   3.89 $\pm$   0.10 &  -0.06 $\pm$   0.05 &   1.16 $\pm$   0.02 & (2) &   3.93 $\pm$   0.23 &   3.82 $\pm$   0.22 \\
\object{KIC8738809} & 6207 $\pm$ 68 &   4.17 $\pm$   0.11 &   0.12 $\pm$   0.06 &   1.65 $\pm$   0.03 & (2) &   3.92 $\pm$   0.25 &   3.86 $\pm$   0.23 \\
\object{KIC8938364} & 5808 $\pm$ 71 &   4.31 $\pm$   0.12 &  -0.10 $\pm$   0.06 &   1.10 $\pm$   0.05 & (2) &   4.24 $\pm$   0.24 &   4.15 $\pm$   0.23 \\
\object{KIC9139151} & 6213 $\pm$ 67 &   4.64 $\pm$   0.11 &   0.17 $\pm$   0.06 &   1.24 $\pm$   0.04 & (2) &   4.39 $\pm$   0.25 &   4.32 $\pm$   0.23 \\
\object{KIC9139163}\_esp & 6577 $\pm$ 69 &   4.44 $\pm$   0.10 &   0.21 $\pm$   0.06 &   1.68 $\pm$   0.04 & (2) &   4.02 $\pm$   0.25 &   3.98 $\pm$   0.23 \\
\object{KIC9139163}\_nar & 6584 $\pm$ 67 &   4.47 $\pm$   0.11 &   0.19 $\pm$   0.05 &   1.70 $\pm$   0.03 & (2) &   4.05 $\pm$   0.25 &   4.01 $\pm$   0.24 \\
\object{KIC9206432} & 6772 $\pm$ 73 &   4.61 $\pm$   0.11 &   0.28 $\pm$   0.06 &   1.92 $\pm$   0.05 & (2) &   4.10 $\pm$   0.25 &   4.08 $\pm$   0.24 \\
\object{KIC9512063} & 5842 $\pm$ 72 &   3.87 $\pm$   0.11 &  -0.15 $\pm$   0.06 &   1.12 $\pm$   0.04 & (2) &   3.79 $\pm$   0.24 &   3.70 $\pm$   0.23 \\
\object{KIC9702369} & 6441 $\pm$ 78 &   4.54 $\pm$   0.11 &   0.14 $\pm$   0.06 &   1.39 $\pm$   0.05 & (2) &   4.18 $\pm$   0.25 &   4.14 $\pm$   0.23 \\
\object{KIC9812850} & 6790 $\pm$ 118 &   4.92 $\pm$   0.13 &  -0.04 $\pm$   0.08 &   2.70 $\pm$   0.27 & (2) &   4.40 $\pm$   0.27 &   4.38 $\pm$   0.25 \\
\object{KIC9955598} & 5380 $\pm$ 68 &   4.33 $\pm$   0.12 &   0.04 $\pm$   0.06 &   0.80 $\pm$   0.06 & (2) &   4.46 $\pm$   0.24 &   4.34 $\pm$   0.22 \\
\object{KIC10018963} & 6354 $\pm$ 69 &   4.32 $\pm$   0.11 &  -0.16 $\pm$   0.05 &   1.79 $\pm$   0.05 & (2) &   4.00 $\pm$   0.25 &   3.95 $\pm$   0.23 \\
\object{KIC10068307} & 6288 $\pm$ 68 &   4.28 $\pm$   0.10 &  -0.11 $\pm$   0.06 &   1.68 $\pm$   0.04 & (2) &   3.99 $\pm$   0.24 &   3.93 $\pm$   0.23 \\
\object{KIC10079226} & 6045 $\pm$ 68 &   4.49 $\pm$   0.11 &   0.17 $\pm$   0.06 &   1.17 $\pm$   0.04 & (2) &   4.31 $\pm$   0.24 &   4.24 $\pm$   0.23 \\
\object{KIC10162436} & 6423 $\pm$ 71 &   4.43 $\pm$   0.11 &   0.01 $\pm$   0.06 &   1.75 $\pm$   0.05 & (2) &   4.08 $\pm$   0.25 &   4.03 $\pm$   0.23 \\
\object{KIC10355856} & 6612 $\pm$ 79 &   4.38 $\pm$   0.11 &  -0.01 $\pm$   0.06 &   1.84 $\pm$   0.05 & (2) &   3.94 $\pm$   0.25 &   3.91 $\pm$   0.24 \\
\object{KIC10454113} & 6216 $\pm$ 68 &   4.46 $\pm$   0.10 &   0.00 $\pm$   0.05 &   1.30 $\pm$   0.04 & (2) &   4.20 $\pm$   0.24 &   4.14 $\pm$   0.23 \\
\object{KIC10462940} & 6268 $\pm$ 68 &   4.48 $\pm$   0.10 &   0.18 $\pm$   0.05 &   1.35 $\pm$   0.03 & (2) &   4.20 $\pm$   0.24 &   4.14 $\pm$   0.23 \\
\object{KIC10516096} & 6094 $\pm$ 70 &   4.47 $\pm$   0.11 &  -0.03 $\pm$   0.06 &   1.39 $\pm$   0.05 & (2) &   4.27 $\pm$   0.24 &   4.20 $\pm$   0.23 \\
\object{KIC10644253} & 6132 $\pm$ 65 &   4.54 $\pm$   0.11 &   0.15 $\pm$   0.05 &   1.21 $\pm$   0.03 & (2) &   4.32 $\pm$   0.24 &   4.26 $\pm$   0.23 \\
\object{KIC11026764} & 5802 $\pm$ 68 &   4.12 $\pm$   0.11 &   0.11 $\pm$   0.06 &   1.30 $\pm$   0.04 & (2) &   4.05 $\pm$   0.24 &   3.96 $\pm$   0.23 \\
\object{KIC11137075} & 5610 $\pm$ 71 &   4.10 $\pm$   0.12 &  -0.06 $\pm$   0.06 &   1.10 $\pm$   0.04 & (2) &   4.12 $\pm$   0.24 &   4.02 $\pm$   0.23 \\
\object{KIC11244118} & 5770 $\pm$ 67 &   4.14 $\pm$   0.11 &   0.35 $\pm$   0.06 &   1.19 $\pm$   0.03 & (2) &   4.09 $\pm$   0.24 &   4.00 $\pm$   0.22 \\
\object{KIC11414712} & 5725 $\pm$ 61 &   3.99 $\pm$   0.10 &  -0.02 $\pm$   0.05 &   1.27 $\pm$   0.01 & (2) &   3.96 $\pm$   0.23 &   3.86 $\pm$   0.22 \\
\object{KIC11717120}\_fies & 5118 $\pm$ 67 &   3.80 $\pm$   0.12 &  -0.27 $\pm$   0.06 &   0.89 $\pm$   0.04 & (2) &   4.05 $\pm$   0.23 &   3.91 $\pm$   0.22 \\
\object{KIC11717120}\_nar & 5137 $\pm$ 65 &   3.87 $\pm$   0.12 &  -0.28 $\pm$   0.05 &   0.83 $\pm$   0.04 & (2) &   4.11 $\pm$   0.23 &   3.97 $\pm$   0.22 \\
\object{KIC12009504} & 6267 $\pm$ 71 &   4.37 $\pm$   0.11 &  -0.03 $\pm$   0.06 &   1.59 $\pm$   0.06 & (2) &   4.09 $\pm$   0.25 &   4.03 $\pm$   0.23 \\
\object{KIC12258514} & 6099 $\pm$ 66 &   4.32 $\pm$   0.10 &   0.10 $\pm$   0.05 &   1.36 $\pm$   0.03 & (2) &   4.12 $\pm$   0.24 &   4.05 $\pm$   0.22 \\
\object{KIC12508433} & 5281 $\pm$ 76 &   3.85 $\pm$   0.13 &   0.21 $\pm$   0.06 &   0.98 $\pm$   0.06 & (2) &   4.02 $\pm$   0.24 &   3.90 $\pm$   0.23 \\
\object{$\beta$Hyi} & 5837 $\pm$ 30 &   4.00 $\pm$   0.12 &  -0.08 $\pm$   0.04 &   1.36 $\pm$   0.05 & (3) &   3.92 $\pm$   0.24 &   3.83 $\pm$   0.23 \\
\object{$\tau$Cet} & 5310 $\pm$ 17 &   4.44 $\pm$   0.03 &  -0.52 $\pm$   0.01 &   0.55 $\pm$   0.04 & (4) &   4.60 $\pm$   0.20 &   4.48 $\pm$   0.19 \\
\object{$\iota$Hor} & 6227 $\pm$ 26 &   4.53 $\pm$   0.06 &   0.19 $\pm$   0.02 &   1.29 $\pm$   0.03 & (4) &   4.27 $\pm$   0.23 &   4.21 $\pm$   0.21 \\
\object{$\delta$Eri} & 5027 $\pm$ 48 &   3.66 $\pm$   0.10 &   0.07 $\pm$   0.03 &   0.93 $\pm$   0.06 & (5) &   3.95 $\pm$   0.22 &   3.81 $\pm$   0.21 \\
\object{ProcyonA} & 6738 $\pm$ 43 &   4.18 $\pm$   0.08 &   0.06 $\pm$   0.03 &   2.08 $\pm$   0.06 & (6) &   3.69 $\pm$   0.24 &   3.66 $\pm$   0.23 \\
\object{$\beta$Vir} & 6217 $\pm$ 31 &   4.28 $\pm$   0.03 &   0.20 $\pm$   0.02 &   1.47 $\pm$   0.04 & (6) &   4.02 $\pm$   0.22 &   3.96 $\pm$   0.20 \\
\object{$\alpha$CenA} & 5844 $\pm$ 42 &   4.30 $\pm$   0.19 &   0.28 $\pm$   0.06 &   1.18 $\pm$   0.05 & (3) &   4.21 $\pm$   0.28 &   4.13 $\pm$   0.27 \\
\object{$\alpha$CenB} & 5234 $\pm$ 63 &   4.40 $\pm$   0.11 &   0.16 $\pm$   0.04 &   0.90 $\pm$   0.12 & (7) &   4.59 $\pm$   0.23 &   4.46 $\pm$   0.22 \\
\object{HR5803} & 6452 $\pm$ 35 &   4.50 $\pm$   0.05 &   0.07 $\pm$   0.02 &   1.68 $\pm$   0.05 & (6) &   4.14 $\pm$   0.23 &   4.09 $\pm$   0.21 \\
\object{$\mu$Ara} & 5798 $\pm$ 33 &   4.31 $\pm$   0.08 &   0.32 $\pm$   0.04 &   1.19 $\pm$   0.04 & (3) &   4.25 $\pm$   0.23 &   4.16 $\pm$   0.21 \\
\object{70OphA} & 5346 $\pm$ 45 &   4.47 $\pm$   0.08 &   0.02 $\pm$   0.03 &   1.03 $\pm$   0.07 & (6) &   4.61 $\pm$   0.22 &   4.49 $\pm$   0.20 \\
\object{$\gamma$Pav} & 6217 $\pm$ 34 &   4.64 $\pm$   0.04 &  -0.62 $\pm$   0.02 &   1.65 $\pm$   0.07 & (6) &   4.38 $\pm$   0.22 &   4.32 $\pm$   0.21 \\

\end{longtable}
\tablebib{(1)~\citet{ME13c}; 
(2) \citet{Mol13}; 
(3) \citet{San05};
(4) \citet{Sou08};
(5) \citet{Tsa13};
(6) This work;
(7) \citet{San13}. 
}

\end{longtab}

\begin{longtab}
\begin{longtable}{cccccc}
\caption{\label{TabParFix} Stellar spectroscopic parameters where the surface gravity was fixed to either the value from the photometric transit light curve or a value obtained through asteroseismology. }\\
\hline\hline
Name & T$_{eff,fix}$ & $\log g_{fix}$ & [Fe/H]$_{fix}$ & $\xi_{fix}$  & Method\\
 & (K) & (dex) & (dex) & (km s$^{-1}$) &  \\
\hline
\endfirsthead
\caption{continued.}\\
\hline\hline
Name & T$_{eff,fix}$ & $\log g_{fix}$ & [Fe/H]$_{fix}$ & $\xi_{fix}$  & Method\\
 & (K) & (dex) & (dex) & (km s$^{-1}$) &  \\
\hline
\endhead
\hline
\endfoot
CoRoT-1 & 6576 $\pm$ 54 &   4.35 $\pm$   0.01 &   0.12 $\pm$   0.04 &   1.58 $\pm$   0.09 & Transit \\
CoRoT-10 & 4823 $\pm$ 155 &   4.61 $\pm$   0.02 &   0.15 $\pm$   0.09 &   0.30 $\pm$   0.34 & Transit \\
CoRoT-12 & 5813 $\pm$ 208 &   4.41 $\pm$   0.02 &   0.22 $\pm$   0.14 &   1.23 $\pm$   0.31 & Transit \\
CoRoT-2 & 5794 $\pm$ 97 &   4.52 $\pm$   0.01 &  -0.05 $\pm$   0.07 &   1.88 $\pm$   0.16 & Transit \\
CoRoT-4 & 6454 $\pm$ 93 &   4.37 $\pm$   0.02 &   0.20 $\pm$   0.06 &   1.99 $\pm$   0.14 & Transit \\
CoRoT-5 & 6253 $\pm$ 70 &   4.41 $\pm$   0.03 &   0.05 $\pm$   0.05 &   1.31 $\pm$   0.09 & Transit \\
CoRoT-7 & 5166 $\pm$ 27 &   4.51 $\pm$   0.02 &   0.01 $\pm$   0.02 &   0.57 $\pm$   0.05 & Transit \\
CoRoT-8 & 5105 $\pm$ 178 &   4.49 $\pm$   0.03 &   0.21 $\pm$   0.11 &   0.54 $\pm$   0.40 & Transit \\
CoRoT-9 & 5524 $\pm$ 36 &   4.47 $\pm$   0.04 &  -0.05 $\pm$   0.03 &   0.64 $\pm$   0.05 & Transit \\
HAT-P-1 & 6176 $\pm$ 27 &   4.40 $\pm$   0.01 &   0.27 $\pm$   0.03 &   1.28 $\pm$   0.05 & Transit \\
HAT-P-11 &  &   4.40 $\pm$   0.01 &  &  & Transit \\
HAT-P-17 & 5171 $\pm$ 55 &   4.52 $\pm$   0.02 &   0.04 $\pm$   0.03 &   0.12 $\pm$   0.10 & Transit \\
HAT-P-20 &  &   4.52 $\pm$   0.02 &  &  & Transit \\
HAT-P-26 & 4989 $\pm$ 55 &   4.56 $\pm$   0.02 &  -0.02 $\pm$   0.04 &   0.43 $\pm$   0.16 & Transit \\
HAT-P-27 & 5144 $\pm$ 55 &   4.51 $\pm$   0.03 &   0.32 $\pm$   0.03 &   0.21 $\pm$   0.09 & Transit \\
HAT-P-30 & 6367 $\pm$ 42 &   4.36 $\pm$   0.01 &   0.14 $\pm$   0.03 &   1.48 $\pm$   0.05 & Transit \\
HAT-P-35 & 6226 $\pm$ 45 &   4.22 $\pm$   0.03 &   0.15 $\pm$   0.03 &   1.44 $\pm$   0.06 & Transit \\
HAT-P-4 &  &   4.22 $\pm$   0.03 &  &  & Transit \\
HAT-P-6 & 7107 $\pm$ 111 &   4.20 $\pm$   0.02 &  -0.05 $\pm$   0.11 &   2.17 $\pm$   1.15 & Transit \\
HAT-P-7 & 6671 $\pm$ 61 &   4.04 $\pm$   0.00 &   0.32 $\pm$   0.07 &   1.77 $\pm$   0.14 & Transit \\
HAT-P-8 & 6649 $\pm$ 61 &   4.19 $\pm$   0.03 &   0.14 $\pm$   0.04 &   2.14 $\pm$   0.09 & Transit \\
HD149026 & 6247 $\pm$ 41 &   4.33 $\pm$   0.03 &   0.32 $\pm$   0.05 &   1.53 $\pm$   0.07 & Transit \\
HD17156 & 6173 $\pm$ 29 &   4.21 $\pm$   0.01 &   0.22 $\pm$   0.04 &   1.83 $\pm$   0.05 & Transit \\
HD189733 & 5274 $\pm$ 146 &   4.60 $\pm$   0.01 &   0.04 $\pm$   0.08 &   1.18 $\pm$   0.33 & Transit \\
HD209458 & 6159 $\pm$ 25 &   4.36 $\pm$   0.00 &   0.06 $\pm$   0.02 &   1.30 $\pm$   0.03 & Transit \\
HD80606 & 5741 $\pm$ 72 &   4.42 $\pm$   0.02 &   0.34 $\pm$   0.09 &   1.38 $\pm$   0.09 & Transit \\
HD97658 & 5092 $\pm$ 36 &   4.59 $\pm$   0.01 &  -0.36 $\pm$   0.02 &   0.50 $\pm$   0.08 & Transit \\
Kepler-17 &  &   4.59 $\pm$   0.01 &  &  & Transit \\
Kepler-21 & 6444 $\pm$ 44 &   4.03 $\pm$   0.05 &  -0.01 $\pm$   0.03 &   1.96 $\pm$   0.07 & Transit \\
KOI-135 &  &   4.03 $\pm$   0.05 &  &  & Transit \\
KOI-204 &  &   4.03 $\pm$   0.05 &  &  & Transit \\
OGLE-TR-10 & 6207 $\pm$ 86 &   4.18 $\pm$   0.04 &   0.25 $\pm$   0.10 &   1.87 $\pm$   0.14 & Transit \\
OGLE-TR-111 & 4700 $\pm$ 177 &   4.54 $\pm$   0.01 &   0.17 $\pm$   0.15 &   0.24 $\pm$   1.38 & Transit \\
OGLE-TR-113 & 4458 $\pm$ 166 &   4.56 $\pm$   0.00 &   0.24 $\pm$   0.06 &   0.07 $\pm$   0.29 & Transit \\
OGLE-TR-132 & 6194 $\pm$ 59 &   4.30 $\pm$   0.03 &   0.34 $\pm$   0.07 &   1.50 $\pm$   0.09 & Transit \\
OGLE-TR-182 & 6108 $\pm$ 64 &   4.15 $\pm$   0.07 &   0.38 $\pm$   0.08 &   1.68 $\pm$   0.09 & Transit \\
OGLE-TR-211 & 6398 $\pm$ 91 &   4.17 $\pm$   0.05 &   0.10 $\pm$   0.10 &   2.48 $\pm$   0.21 & Transit \\
OGLE-TR-56 & 6103 $\pm$ 62 &   4.09 $\pm$   0.01 &   0.31 $\pm$   0.08 &   1.38 $\pm$   0.11 & Transit \\
TrES-1 &  &   4.09 $\pm$   0.01 &  &  & Transit \\
TrES-2 &  &   4.09 $\pm$   0.01 &  &  & Transit \\
TrES-3 &  &   4.09 $\pm$   0.01 &  &  & Transit \\
TrES-4 &  &   4.09 $\pm$   0.01 &  &  & Transit \\
WASP-1 & 6270 $\pm$ 45 &   4.23 $\pm$   0.03 &   0.24 $\pm$   0.03 &   1.46 $\pm$   0.05 & Transit \\
WASP-10 & 4553 $\pm$ 125 &   4.61 $\pm$   0.02 &   0.07 $\pm$   0.05 &   0.11 $\pm$   0.47 & Transit \\
WASP-11 & 4789 $\pm$ 125 &   4.63 $\pm$   0.02 &   0.05 $\pm$   0.05 &   0.27 $\pm$   0.24 & Transit \\
WASP-12 & 6365 $\pm$ 52 &   4.05 $\pm$   0.02 &   0.24 $\pm$   0.04 &   1.76 $\pm$   0.07 & Transit \\
WASP-13 & 6127 $\pm$ 21 &   3.89 $\pm$   0.03 &   0.13 $\pm$   0.05 &   1.46 $\pm$   0.10 & Transit \\
WASP-15 & 6692 $\pm$ 70 &   4.22 $\pm$   0.02 &   0.16 $\pm$   0.04 &   1.96 $\pm$   0.09 & Transit \\
WASP-16 & 5617 $\pm$ 22 &   4.49 $\pm$   0.02 &   0.09 $\pm$   0.02 &   0.70 $\pm$   0.03 & Transit \\
WASP-17 & 6822 $\pm$ 83 &   4.16 $\pm$   0.01 &  -0.09 $\pm$   0.05 &   2.67 $\pm$   0.22 & Transit \\
WASP-18 & 6603 $\pm$ 69 &   4.32 $\pm$   0.03 &   0.24 $\pm$   0.05 &   2.00 $\pm$   0.10 & Transit \\
WASP-19 & 5612 $\pm$ 62 &   4.44 $\pm$   0.01 &   0.27 $\pm$   0.05 &   1.27 $\pm$   0.09 & Transit \\
WASP-2 & 5047 $\pm$ 72 &   4.54 $\pm$   0.01 &   0.00 $\pm$   0.05 &   0.48 $\pm$   0.12 & Transit \\
WASP-21 & 5961 $\pm$ 55 &   4.28 $\pm$   0.03 &  -0.20 $\pm$   0.04 &   1.16 $\pm$   0.08 & Transit \\
WASP-22 & 6230 $\pm$ 46 &   4.32 $\pm$   0.02 &   0.30 $\pm$   0.03 &   1.51 $\pm$   0.06 & Transit \\
WASP-23 & 4965 $\pm$ 99 &   4.59 $\pm$   0.01 &   0.03 $\pm$   0.06 &   0.41 $\pm$   0.23 & Transit \\
WASP-24 & 6426 $\pm$ 58 &   4.25 $\pm$   0.01 &   0.17 $\pm$   0.04 &   1.69 $\pm$   0.08 & Transit \\
WASP-25 & 5741 $\pm$ 35 &   4.51 $\pm$   0.01 &   0.06 $\pm$   0.03 &   1.12 $\pm$   0.05 & Transit \\
WASP-26 & 6084 $\pm$ 31 &   4.25 $\pm$   0.03 &   0.19 $\pm$   0.02 &   1.40 $\pm$   0.04 & Transit \\
WASP-28 & 6161 $\pm$ 38 &   4.44 $\pm$   0.03 &  -0.11 $\pm$   0.03 &   1.25 $\pm$   0.06 & Transit \\
WASP-29 &  &   4.44 $\pm$   0.03 &  &  & Transit \\
WASP-31 & 6524 $\pm$ 75 &   4.31 $\pm$   0.02 &  -0.03 $\pm$   0.05 &   1.83 $\pm$   0.11 & Transit \\
WASP-32 & 6410 $\pm$ 141 &   4.32 $\pm$   0.03 &   0.14 $\pm$   0.10 &   1.25 $\pm$   0.21 & Transit \\
WASP-34 & 5691 $\pm$ 26 &   4.37 $\pm$   0.05 &   0.07 $\pm$   0.02 &   0.94 $\pm$   0.03 & Transit \\
WASP-35 & 6167 $\pm$ 62 &   4.39 $\pm$   0.02 &   0.01 $\pm$   0.05 &   1.49 $\pm$   0.09 & Transit \\
WASP-36 & 5939 $\pm$ 59 &   4.49 $\pm$   0.01 &  -0.01 $\pm$   0.05 &   0.92 $\pm$   0.09 & Transit \\
WASP-38 & 6510 $\pm$ 60 &   4.27 $\pm$   0.01 &   0.11 $\pm$   0.04 &   1.95 $\pm$   0.09 & Transit \\
WASP-4 & 5518 $\pm$ 43 &   4.49 $\pm$   0.01 &   0.03 $\pm$   0.03 &   0.87 $\pm$   0.07 & Transit \\
WASP-41 & 5573 $\pm$ 33 &   4.49 $\pm$   0.03 &   0.07 $\pm$   0.02 &   1.15 $\pm$   0.05 & Transit \\
WASP-42 & 5030 $\pm$ 79 &   4.52 $\pm$   0.02 &   0.31 $\pm$   0.05 &   0.48 $\pm$   0.13 & Transit \\
WASP-47 & 5536 $\pm$ 68 &   4.34 $\pm$   0.01 &   0.36 $\pm$   0.05 &   1.17 $\pm$   0.09 & Transit \\
WASP-5 & 5839 $\pm$ 83 &   4.39 $\pm$   0.03 &   0.20 $\pm$   0.06 &   1.06 $\pm$   0.12 & Transit \\
WASP-50 & 5459 $\pm$ 42 &   4.48 $\pm$   0.02 &   0.11 $\pm$   0.03 &   1.12 $\pm$   0.06 & Transit \\
WASP-54 & 6361 $\pm$ 40 &   4.00 $\pm$   0.02 &   0.04 $\pm$   0.03 &   1.59 $\pm$   0.05 & Transit \\
WASP-55 & 6145 $\pm$ 53 &   4.41 $\pm$   0.01 &   0.14 $\pm$   0.04 &   1.24 $\pm$   0.06 & Transit \\
WASP-6 & 5392 $\pm$ 41 &   4.52 $\pm$   0.00 &  -0.13 $\pm$   0.03 &   0.82 $\pm$   0.07 & Transit \\
WASP-62 & 6511 $\pm$ 70 &   4.33 $\pm$   0.02 &   0.31 $\pm$   0.05 &   1.74 $\pm$   0.09 & Transit \\
WASP-63 & 5832 $\pm$ 60 &   4.00 $\pm$   0.02 &   0.33 $\pm$   0.05 &   1.49 $\pm$   0.07 & Transit \\
WASP-66 & 7079 $\pm$ 79 &   4.10 $\pm$   0.03 &   0.10 $\pm$   0.05 &   3.14 $\pm$   0.27 & Transit \\
WASP-67 & 5220 $\pm$ 85 &   4.51 $\pm$   0.02 &   0.15 $\pm$   0.06 &   0.75 $\pm$   0.12 & Transit \\
WASP-7 & 6638 $\pm$ 155 &   4.22 $\pm$   0.04 &   0.14 $\pm$   0.09 &   3.09 $\pm$   0.83 & Transit \\
WASP-71 & 6215 $\pm$ 52 &   3.92 $\pm$   0.03 &   0.39 $\pm$   0.04 &   1.75 $\pm$   0.06 & Transit \\
WASP-77A & 5503 $\pm$ 41 &   4.48 $\pm$   0.00 &   0.03 $\pm$   0.03 &   0.84 $\pm$   0.06 & Transit \\
WASP-78 & 6317 $\pm$ 71 &   3.89 $\pm$   0.03 &  -0.05 $\pm$   0.05 &   1.71 $\pm$   0.10 & Transit \\
WASP-79 & 7052 $\pm$ 162 &   4.07 $\pm$   0.03 &   0.25 $\pm$   0.10 &   2.72 $\pm$   0.24 & Transit \\
WASP-8 & 5648 $\pm$ 36 &   4.48 $\pm$   0.01 &   0.28 $\pm$   0.03 &   1.15 $\pm$   0.05 & Transit \\
XO-1 & 5922 $\pm$ 42 &   4.50 $\pm$   0.01 &   0.03 $\pm$   0.05 &   1.14 $\pm$   0.09 & Transit \\
KIC1430163 & 6879 $\pm$ 59 &   4.22 $\pm$   0.01 &   0.06 $\pm$   0.15 &   2.20 $\pm$   0.11 & Seismic \\
KIC1435467 & 6548 $\pm$ 68 &   4.11 $\pm$   0.01 &   0.12 $\pm$   0.15 &   2.15 $\pm$   0.09 & Seismic \\
KIC3427720 & 6237 $\pm$ 35 &   3.97 $\pm$   0.01 &   0.10 $\pm$   0.11 &   1.53 $\pm$   0.04 & Seismic \\
KIC3456181 & 6621 $\pm$ 74 &   3.83 $\pm$   0.02 &   0.02 $\pm$   0.19 &   2.10 $\pm$   0.11 & Seismic \\
KIC3632418 & 6459 $\pm$ 47 &   3.77 $\pm$   0.01 &   0.01 $\pm$   0.13 &   2.00 $\pm$   0.06 & Seismic \\
KIC3643774 & 6212 $\pm$ 46 &   4.03 $\pm$   0.01 &   0.30 $\pm$   0.10 &   1.55 $\pm$   0.05 & Seismic \\
KIC3656476 & 5784 $\pm$ 24 &   4.13 $\pm$   0.01 &   0.31 $\pm$   0.03 &   1.24 $\pm$   0.03 & Seismic \\
KIC4072740 &  &   3.77 $\pm$   0.01 &  &  & Seismic \\
KIC4346201 & 6274 $\pm$ 68 &   3.95 $\pm$   0.01 &  -0.15 $\pm$   0.12 &   1.74 $\pm$   0.10 & Seismic \\
KIC4586099 & 6560 $\pm$ 53 &   4.05 $\pm$   0.02 &  -0.02 $\pm$   0.09 &   1.91 $\pm$   0.08 & Seismic \\
KIC4638884 & 6700 $\pm$ 78 &   4.03 $\pm$   0.01 &  -0.03 $\pm$   0.30 &   3.46 $\pm$   0.27 & Seismic \\
KIC4914923 & 6038 $\pm$ 28 &   4.04 $\pm$   0.01 &   0.23 $\pm$   0.06 &   1.43 $\pm$   0.03 & Seismic \\
KIC4931390 & 6907 $\pm$ 51 &   3.96 $\pm$   0.02 &   0.03 $\pm$   0.13 &   2.00 $\pm$   0.08 & Seismic \\
KIC5184732\_esp & 5972 $\pm$ 34 &   4.13 $\pm$   0.01 &   0.46 $\pm$   0.05 &   1.32 $\pm$   0.03 & Seismic \\
KIC5184732\_nar & 5971 $\pm$ 34 &   4.13 $\pm$   0.01 &   0.44 $\pm$   0.06 &   1.31 $\pm$   0.04 & Seismic \\
KIC5371516 & 6582 $\pm$ 98 &   3.62 $\pm$   0.01 &   0.17 $\pm$   0.37 &   2.48 $\pm$   0.15 & Seismic \\
KIC5450445 & 6478 $\pm$ 52 &   3.89 $\pm$   0.01 &   0.29 $\pm$   0.14 &   1.91 $\pm$   0.06 & Seismic \\
KIC5512589 & 5860 $\pm$ 31 &   3.81 $\pm$   0.01 &   0.15 $\pm$   0.05 &   1.31 $\pm$   0.03 & Seismic \\
KIC5773345 & 6470 $\pm$ 43 &   3.53 $\pm$   0.01 &   0.36 $\pm$   0.15 &   2.06 $\pm$   0.05 & Seismic \\
KIC5955122 & 6107 $\pm$ 39 &   4.16 $\pm$   0.01 &  -0.05 $\pm$   0.04 &   1.70 $\pm$   0.05 & Seismic \\
KIC6116048 & 6267 $\pm$ 32 &   3.88 $\pm$   0.01 &  -0.07 $\pm$   0.11 &   1.64 $\pm$   0.04 & Seismic \\
KIC6225718 & 6463 $\pm$ 36 &   3.97 $\pm$   0.01 &  -0.01 $\pm$   0.11 &   1.74 $\pm$   0.05 & Seismic \\
KIC6442183 & 5795 $\pm$ 18 &   3.90 $\pm$   0.01 &  -0.09 $\pm$   0.03 &   1.29 $\pm$   0.02 & Seismic \\
KIC6603624 & 5872 $\pm$ 51 &   4.13 $\pm$   0.01 &   0.34 $\pm$   0.13 &   1.46 $\pm$   0.05 & Seismic \\
KIC6933899 & 5983 $\pm$ 28 &   3.81 $\pm$   0.01 &   0.08 $\pm$   0.06 &   1.42 $\pm$   0.03 & Seismic \\
KIC7103006 & 6756 $\pm$ 50 &   4.09 $\pm$   0.01 &   0.24 $\pm$   0.11 &   2.09 $\pm$   0.07 & Seismic \\
KIC7668623 & 6611 $\pm$ 86 &   4.03 $\pm$   0.01 &   0.06 $\pm$   0.25 &   2.63 $\pm$   0.19 & Seismic \\
KIC7680114 & 6061 $\pm$ 30 &   4.01 $\pm$   0.01 &   0.18 $\pm$   0.08 &   1.51 $\pm$   0.04 & Seismic \\
KIC7747078 & 6129 $\pm$ 52 &   4.26 $\pm$   0.01 &  -0.10 $\pm$   0.05 &   1.70 $\pm$   0.07 & Seismic \\
KIC7799349 &  &   3.61 $\pm$   0.01 &  &  & Seismic \\
KIC7940546\_esp & 6462 $\pm$ 55 &   4.14 $\pm$   0.01 &  -0.09 $\pm$   0.11 &   2.19 $\pm$   0.10 & Seismic \\
KIC7940546\_nar & 6507 $\pm$ 57 &   4.14 $\pm$   0.01 &  -0.09 $\pm$   0.15 &   2.43 $\pm$   0.13 & Seismic \\
KIC7976303 & 6193 $\pm$ 49 &   4.22 $\pm$   0.01 &  -0.41 $\pm$   0.03 &   1.59 $\pm$   0.08 & Seismic \\
KIC8006161\_esp & 5688 $\pm$ 53 &   4.14 $\pm$   0.01 &   0.34 $\pm$   0.19 &   1.52 $\pm$   0.07 & Seismic \\
KIC8006161\_nar & 5656 $\pm$ 50 &   4.14 $\pm$   0.01 &   0.32 $\pm$   0.15 &   1.50 $\pm$   0.06 & Seismic \\
KIC8026226 & 6479 $\pm$ 60 &   3.96 $\pm$   0.01 &  -0.12 $\pm$   0.14 &   2.78 $\pm$   0.17 & Seismic \\
KIC8179536 & 6635 $\pm$ 42 &   4.14 $\pm$   0.01 &   0.19 $\pm$   0.10 &   1.79 $\pm$   0.05 & Seismic \\
KIC8228742 & 6318 $\pm$ 49 &   4.25 $\pm$   0.01 &   0.01 $\pm$   0.06 &   1.77 $\pm$   0.07 & Seismic \\
KIC8379927\_esp & 6327 $\pm$ 72 &   4.02 $\pm$   0.01 &  -0.17 $\pm$   0.29 &   2.36 $\pm$   0.13 & Seismic \\
KIC8379927\_nar & 6343 $\pm$ 45 &   4.02 $\pm$   0.01 &  -0.12 $\pm$   0.12 &   1.24 $\pm$   0.05 & Seismic \\
KIC8394589 & 6321 $\pm$ 45 &   3.85 $\pm$   0.02 &  -0.19 $\pm$   0.15 &   1.61 $\pm$   0.06 & Seismic \\
KIC8524425 & 5729 $\pm$ 26 &   3.91 $\pm$   0.01 &   0.16 $\pm$   0.04 &   1.28 $\pm$   0.03 & Seismic \\
KIC8561221 & 5376 $\pm$ 34 &   3.72 $\pm$   0.01 &  -0.04 $\pm$   0.04 &   1.20 $\pm$   0.04 & Seismic \\
KIC8694723\_nar & 6471 $\pm$ 54 &   4.20 $\pm$   0.01 &  -0.38 $\pm$   0.11 &   2.01 $\pm$   0.11 & Seismic \\
KIC8694723\_fies & 6515 $\pm$ 55 &   4.21 $\pm$   0.01 &  -0.33 $\pm$   0.10 &   2.06 $\pm$   0.12 & Seismic \\
KIC8702606 & 5640 $\pm$ 19 &   3.64 $\pm$   0.01 &  -0.03 $\pm$   0.04 &   1.29 $\pm$   0.02 & Seismic \\
KIC8738809 & 6247 $\pm$ 33 &   3.94 $\pm$   0.02 &   0.15 $\pm$   0.05 &   1.72 $\pm$   0.04 & Seismic \\
KIC8938364 & 5891 $\pm$ 44 &   3.99 $\pm$   0.01 &  -0.05 $\pm$   0.09 &   1.30 $\pm$   0.05 & Seismic \\
KIC9139151 & 6351 $\pm$ 33 &   4.10 $\pm$   0.01 &   0.25 $\pm$   0.10 &   1.53 $\pm$   0.03 & Seismic \\
KIC9139163\_esp & 6641 $\pm$ 39 &   4.09 $\pm$   0.02 &   0.25 $\pm$   0.07 &   1.79 $\pm$   0.04 & Seismic \\
KIC9139163\_nar & 6645 $\pm$ 34 &   4.09 $\pm$   0.02 &   0.24 $\pm$   0.06 &   1.81 $\pm$   0.04 & Seismic \\
KIC9206432 & 6838 $\pm$ 47 &   4.00 $\pm$   0.01 &   0.34 $\pm$   0.12 &   2.04 $\pm$   0.05 & Seismic \\
KIC9512063 & 5890 $\pm$ 44 &   3.51 $\pm$   0.01 &  -0.12 $\pm$   0.09 &   1.24 $\pm$   0.05 & Seismic \\
KIC9702369 & 6567 $\pm$ 55 &   3.92 $\pm$   0.01 &   0.22 $\pm$   0.16 &   1.62 $\pm$   0.06 & Seismic \\
KIC9812850 & 6831 $\pm$ 101 &   4.12 $\pm$   0.01 &   0.00 $\pm$   0.45 &   2.84 $\pm$   0.25 & Seismic \\
KIC9955598 & 5593 $\pm$ 38 &   3.98 $\pm$   0.01 &   0.09 $\pm$   0.13 &   1.32 $\pm$   0.04 & Seismic \\
KIC10018963 & 6369 $\pm$ 33 &   4.20 $\pm$   0.01 &  -0.15 $\pm$   0.03 &   1.83 $\pm$   0.06 & Seismic \\
KIC10068307 & 6312 $\pm$ 33 &   4.11 $\pm$   0.02 &  -0.10 $\pm$   0.04 &   1.74 $\pm$   0.05 & Seismic \\
KIC10079226 & 6162 $\pm$ 33 &   4.11 $\pm$   0.01 &   0.23 $\pm$   0.08 &   1.40 $\pm$   0.04 & Seismic \\
KIC10162436 & 6454 $\pm$ 36 &   4.21 $\pm$   0.01 &   0.03 $\pm$   0.05 &   1.82 $\pm$   0.05 & Seismic \\
KIC10355856 & 6672 $\pm$ 46 &   3.86 $\pm$   0.01 &   0.04 $\pm$   0.10 &   1.92 $\pm$   0.06 & Seismic \\
KIC10454113 & 6330 $\pm$ 36 &   3.72 $\pm$   0.01 &   0.07 $\pm$   0.12 &   1.57 $\pm$   0.04 & Seismic \\
KIC10462940 & 6375 $\pm$ 32 &   3.89 $\pm$   0.01 &   0.25 $\pm$   0.09 &   1.56 $\pm$   0.03 & Seismic \\
KIC10516096 & 6182 $\pm$ 37 &   4.01 $\pm$   0.01 &   0.03 $\pm$   0.10 &   1.59 $\pm$   0.05 & Seismic \\
KIC10644253 & 6262 $\pm$ 29 &   4.03 $\pm$   0.01 &   0.23 $\pm$   0.08 &   1.47 $\pm$   0.03 & Seismic \\
KIC11026764 & 5837 $\pm$ 33 &   3.99 $\pm$   0.01 &   0.12 $\pm$   0.04 &   1.37 $\pm$   0.04 & Seismic \\
KIC11137075 & 5656 $\pm$ 36 &   3.98 $\pm$   0.01 &  -0.04 $\pm$   0.05 &   1.20 $\pm$   0.04 & Seismic \\
KIC11244118 & 5816 $\pm$ 31 &   4.01 $\pm$   0.01 &   0.37 $\pm$   0.04 &   1.28 $\pm$   0.03 & Seismic \\
KIC11414712 & 5694 $\pm$ 15 &   4.07 $\pm$   0.01 &  -0.04 $\pm$   0.01 &   1.21 $\pm$   0.02 & Seismic \\
KIC11717120\_fies & 5167 $\pm$ 30 &   3.70 $\pm$   0.01 &  -0.26 $\pm$   0.04 &   1.01 $\pm$   0.04 & Seismic \\
KIC11717120\_nar & 5211 $\pm$ 25 &   3.70 $\pm$   0.01 &  -0.27 $\pm$   0.04 &   1.04 $\pm$   0.03 & Seismic \\
KIC12009504 & 6322 $\pm$ 38 &   3.91 $\pm$   0.02 &   0.01 $\pm$   0.10 &   1.76 $\pm$   0.06 & Seismic \\
KIC12258514 & 6153 $\pm$ 29 &   4.10 $\pm$   0.02 &   0.13 $\pm$   0.05 &   1.48 $\pm$   0.03 & Seismic \\
KIC12508433 & 5313 $\pm$ 48 &   3.79 $\pm$   0.01 &   0.21 $\pm$   0.05 &   1.06 $\pm$   0.06 & Seismic \\
$\beta$Hyi &  &   3.96 $\pm$   0.01 &  &  & Seismic \\
$\tau$Cet &  &   4.57 $\pm$   0.00 &  &  & Seismic \\
$\iota$Hor &  &   4.38 $\pm$   0.01 &  &  & Seismic \\
$\delta$Eri &  &   3.79 $\pm$   0.01 &  &  & Seismic \\
ProcyonA & 6749 $\pm$ 43 &   3.98 $\pm$   0.01 &   0.07 $\pm$   0.05 &   2.10 $\pm$   0.06 & Seismic \\
$\beta$Vir & 6251 $\pm$ 31 &   4.12 $\pm$   0.01 &   0.22 $\pm$   0.04 &   1.54 $\pm$   0.03 & Seismic \\
$\alpha$CenA &  &   4.32 $\pm$   0.01 &  &  & Seismic \\
$\alpha$CenB &  &   4.53 $\pm$   0.01 &  &  & Seismic \\
HR5803 & 6498 $\pm$ 34 &   4.21 $\pm$   0.01 &   0.10 $\pm$   0.06 &   1.78 $\pm$   0.05 & Seismic \\
$\mu$Ara &  &   4.23 $\pm$   0.01 &  &  & Seismic \\
70OphA & 5218 $\pm$ 40 &   4.54 $\pm$   0.01 &   0.02 $\pm$   0.11 &   0.61 $\pm$   0.11 & Seismic \\
$\gamma$Pav & 6253 $\pm$ 32 &   4.36 $\pm$   0.01 &  -0.60 $\pm$   0.06 &   1.79 $\pm$   0.08 & Seismic \\

\end{longtable}
\end{longtab}

\section{Surface gravity from transits}\label{Tra}

For stars with transiting planets, an independent measurement of the surface gravity can be obtained using the effective temperature and metallicity from the spectroscopic analysis, and the stellar density which is obtained directly from the transit light curve through the formula

\begin{equation}
\rho_{\ast} + k^3\rho_p = \frac{3\pi}{\mathrm{G}P^2}\left(\frac{a}{R_{\ast}}\right)^3,
\end{equation}

\noindent where $\rho_{\ast}$ and $\rho_p$ are the stellar and planetary density, $P$ the period of the planet, $a$ the orbital separation, G the gravitational constant, and $R_{\ast}$ the stellar radius \citep{Winn11}. Since the constant coefficient $k$ is usually small, the second term on the left is negligible. All parameters on the right come directly from analysing the transit light curve.

The surface gravities can then be obtained through isochrone fitting using the PARSEC isochrones \citep{Bre12} and a $\chi^2$ minimization process \citep[for details, see ][]{ME13c}. They showed that the spectroscopic and photometric surface gravities do not compare well with each other. The $\log g$ values obtained through the photometric transit light curve compare best with literature values (but note that most literate values also come from photometric methods).

In this work, we used the sample of 87 stars from \citet{ME13c}. All these stars are of spectral type F, G or K and are known to be orbited by a transiting planet (according to the online catalog \url{www.exoplanet.eu}). They were observed with different high-resolution spectrographs and analysed in \citet{ME13c} with our method (see Table \ref{TabPar}).

In order to test the effect the surface gravity has on the determination of the other three atmospheric parameters, we redid the same spectroscopic analysis as performed in \citet{ME13c}, but we fixed the surface gravity to the value obtained through the photometric transit light curve. The results can be found in Table \ref{TabParFix}. The errors of the effective temperature, metallicity and microturbulence were set to the errors of the unconstrained values. Not all spectra were suitable to derive atmospheric parameters whilst fixing one parameter due to their lower signal-to-noise ratio (S/N). For these lower S/N stars we did not always reach the rigorous convergence we apply in the analysis and we preferred not to lighten it. In the end, we got results for 76 out of the 87 stars. This subsample is representable for the complete sample.

For 12 of the cooler stars, where the shorter linelist of \citet{Tsa13} was used, we did not always converge to a good microturbulence determination because of the small EW interval of the measured \ion{Fe}{i} lines. Following \citet{ME13b}, the microturbulence was derived with the empirical formula \citep[taken from ][]{Ram13}

\begin{multline}\label{EqVt}
\xi_t = 1.163 + 7.808\cdot10^{-4}\cdot (T_{\mathrm{eff}}-5800) \\- 0.494 \cdot (\log g - 4.30) - 0.05\cdot [Fe/H].
\end{multline}

This formula is comparable to what \citet{Tsa13} found, using 451 FGK dwarfs with parameters derived following our method. In this work, however, we gave preference to the formula of \citet{Ram13}, since they include the metallicity of the star in the relation.

We compared the stellar parameters obtained from fixing the surface gravity to the photometric light curve value with the parameters obtained with no constraints on the surface gravity \citep[taken from ][]{ME13c}. All three parameters compare well, with mean differences of $19$\,K, $0.02$\,dex and $0.0$\,km/s for the effective temperature, metallicity, and microturbulence, respectively.
In Figure \ref{FigFL1}, the differences in the spectroscopic parameters (defined as `constrained with transit $\log g$ - unconstrained') are plotted against the difference in surface gravity (defined as 'photometric - spectroscopic'). All three parameters are anticorrelated with the difference in surface gravity. 

Because of these trends, we calculated the median absolute deviations (MAD) as well, which is an easy way to quantify variation. We find that the MADs are $66.5$\,K, $0.03$\,dex and $0.13$\,km/s for the effective temperature, metallicity, and microturbulence, respectively. Since these values are within the errorbars of the parameters, these trends are thus small enough so that we are confident that the surface gravity does not have a large effect on the determination of other atmospheric parameters using our method of spectral line analysis with the linelists of \citet{Sou08} and \citet{Tsa13}.

The differences in the spectroscopic parameters become constant for higher absolute differences of the surface gravity. This is in contrast with the results from \citet{Tor12} where the differences were linearly correlated with the surface gravity difference, also for the larger differences. In their work, they used two spectral synthesis methods, SPC \citep[Stellar Parameter
Classification - ][]{Buch12} and SME \citep[Spectroscopy Made Easy - ][]{Val96}. They also tested for a spectral line analysis method, but the sample was too small for any firm conclusions.

\subsection{Correction with temperature}\label{CorrTr}

\begin{figure}[t!]
\begin{center}
\includegraphics[width=7cm]{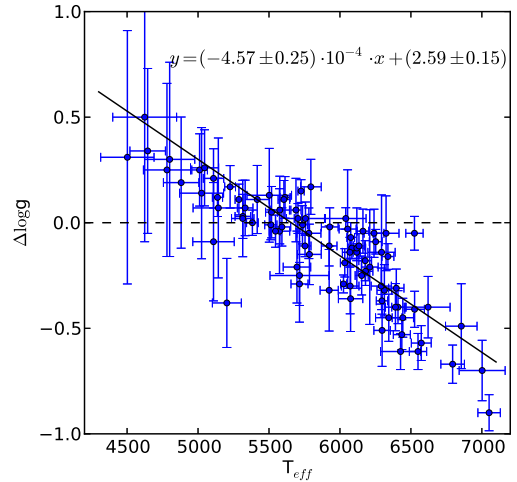}
\caption{Surface gravity difference (`photometric - spectroscopic') versus the (unconstrained) effective temperature. A linear fit is shown with the solid black curve.}
\label{Figlogg}
\end{center}
\end{figure}

The differences in photometric and spectroscopic surface gravity seem to depend on the (unconstrained) effective temperature as can be seen in Figure \ref{Figlogg}, where a decreasing linear trend is noticeable. The same trend is found for the microturbulence, which is closely related to the effective temperature (as seen from Equation \ref{EqVt}). Comparing the $\log$g differences with metallicities reveals no additional trends. 

We fitted the trend with temperature with a linear function, taking into account the errors on both datasets (see Figure \ref{Figlogg}). We used the complete sample of 87 stars and followed the procedure as described in Numerical Recipes in C \citep{Pre92C} to obtain 1-sigma errors on the coefficients. We found the following relation:

\begin{equation}\label{EqFit}
\log g_{LC} - \log g_{spec} = -4.57\pm0.25\cdot 10^{-4} \cdot T_{\mathrm{eff}} + 2.59 \pm 0.15
\end{equation}

This formula is valid for stars with an effective temperature between $4500$\,K and $7050$\,K. It can be used to correct for the spectroscopic surface gravity when no transit light curve is available (and thus even for stars without planets). Using this formula assumes that the $\log g$ value coming from the transit is the more accurate one. As we will show later (see Section \ref{Seis}), these values can also suffer from inaccuracies. By applying this formula, we corrected our spectroscopic surface gravities for the sample of 87 stars (see Table \ref{TabPar}). The resulting values compare, as expected, better with the photometric surface gravities. 

As an additional test, we selected a subsample of our sample of stars, the ones with the highest S/N spectra (38 out of 87 stars). The coolest stars were hereby left out of the sample. We then again redid the spectroscopic analysis, but this time we fixed the surface gravity to the value corrected using Equation \ref{EqFit}..

\begin{figure*}[ht!]
\begin{center}
\includegraphics[width=5.8cm]{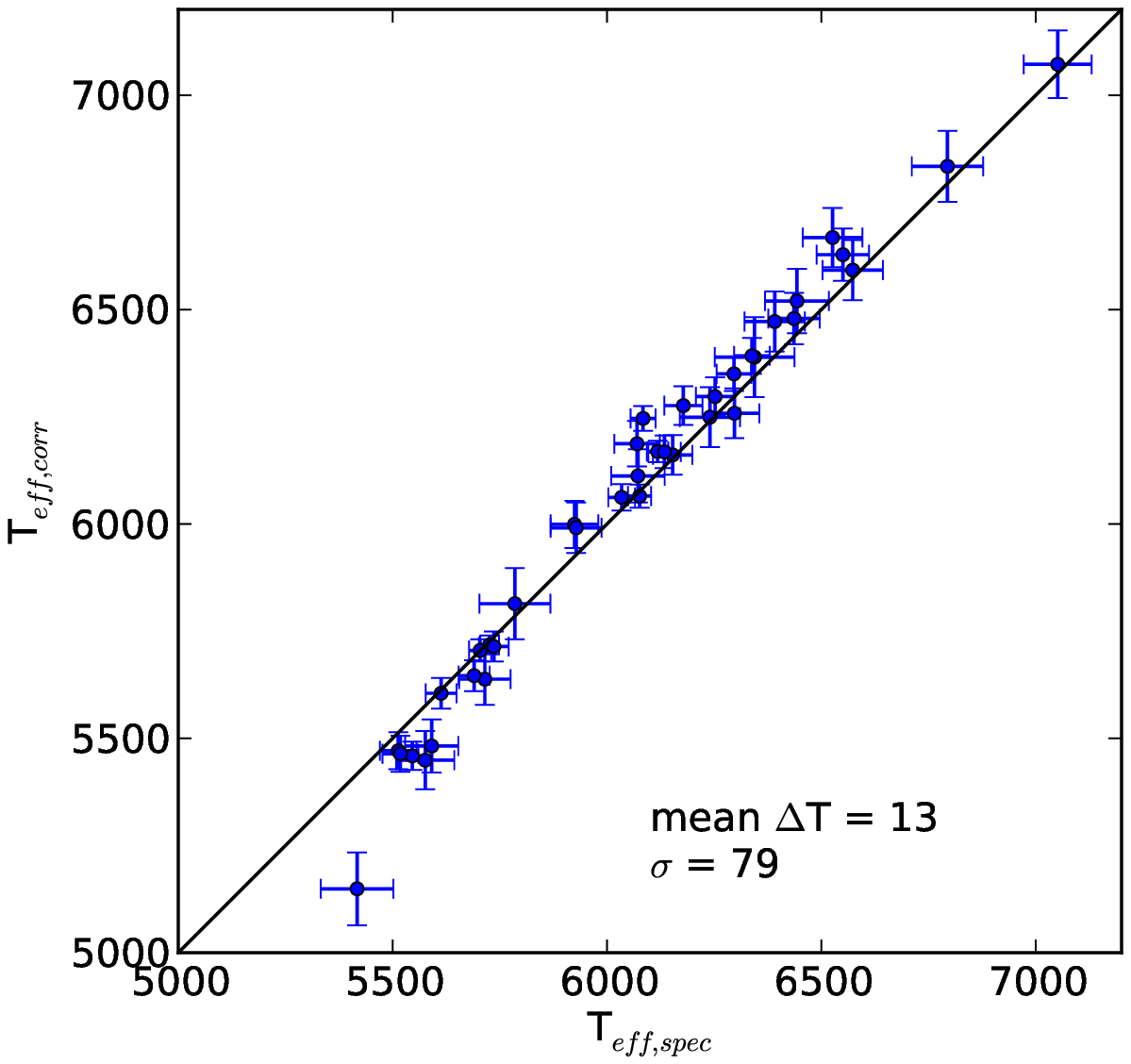}
\includegraphics[width=5.8cm]{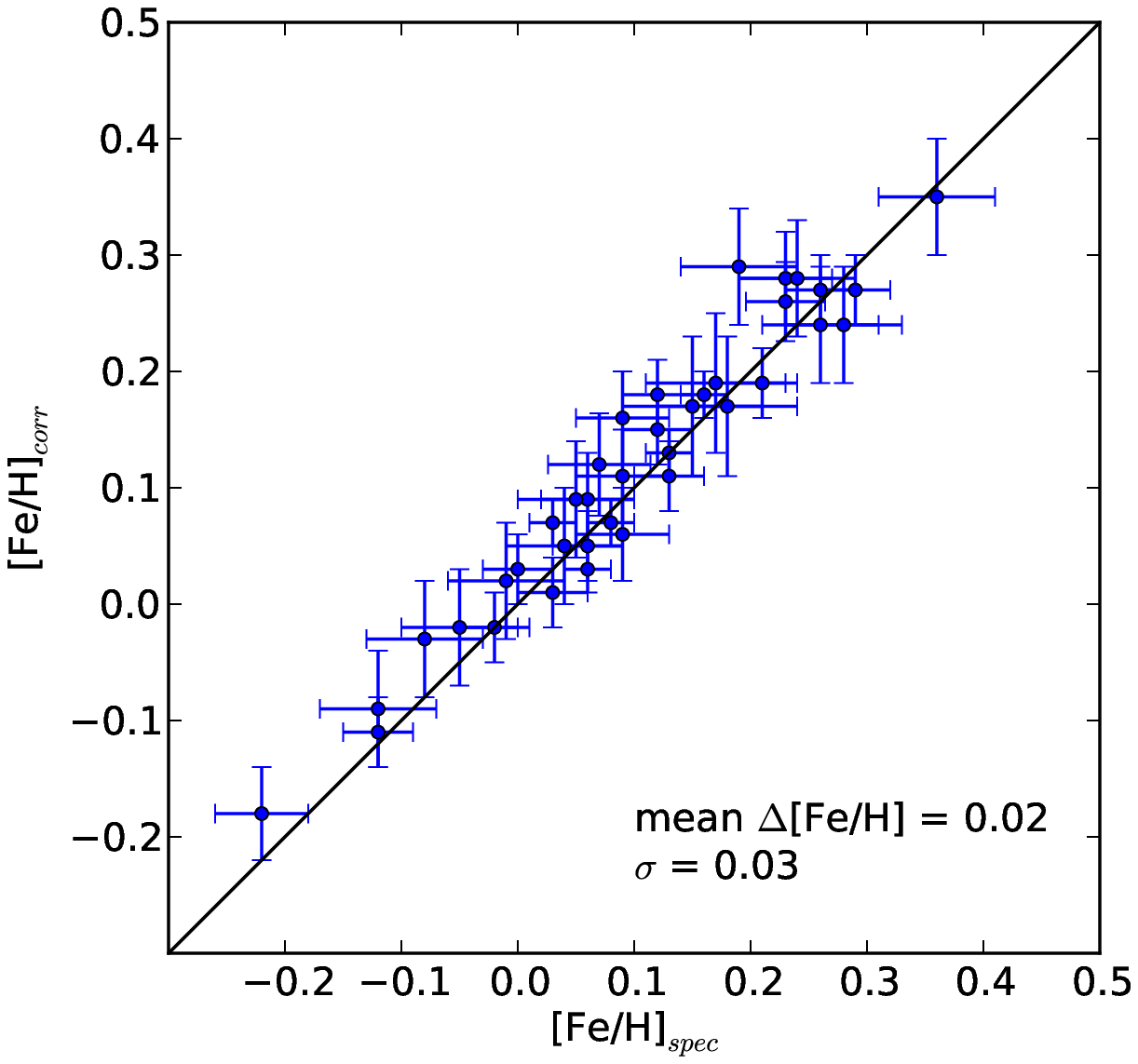}
\includegraphics[width=5.8cm]{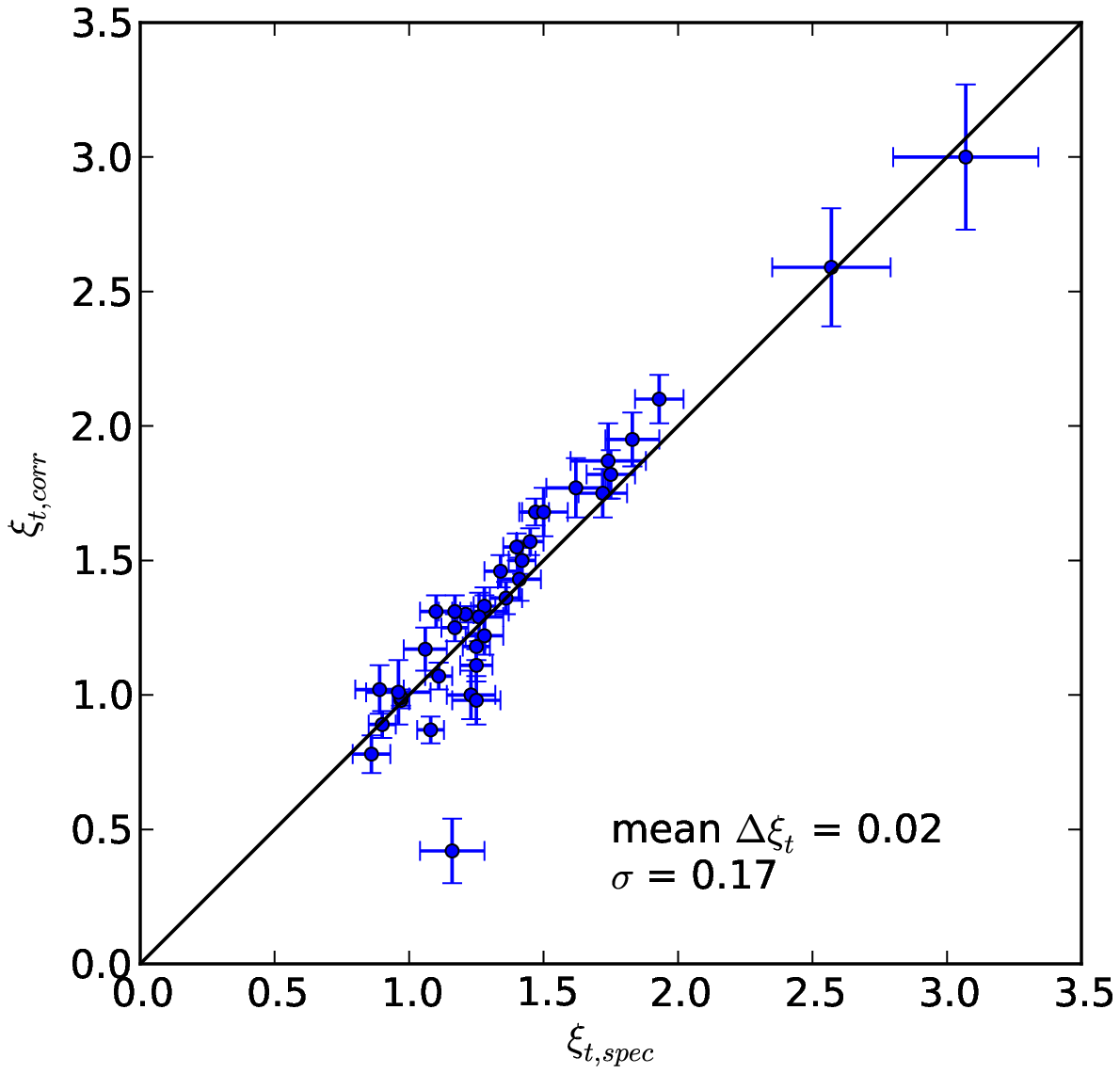}\\
\includegraphics[width=5.8cm]{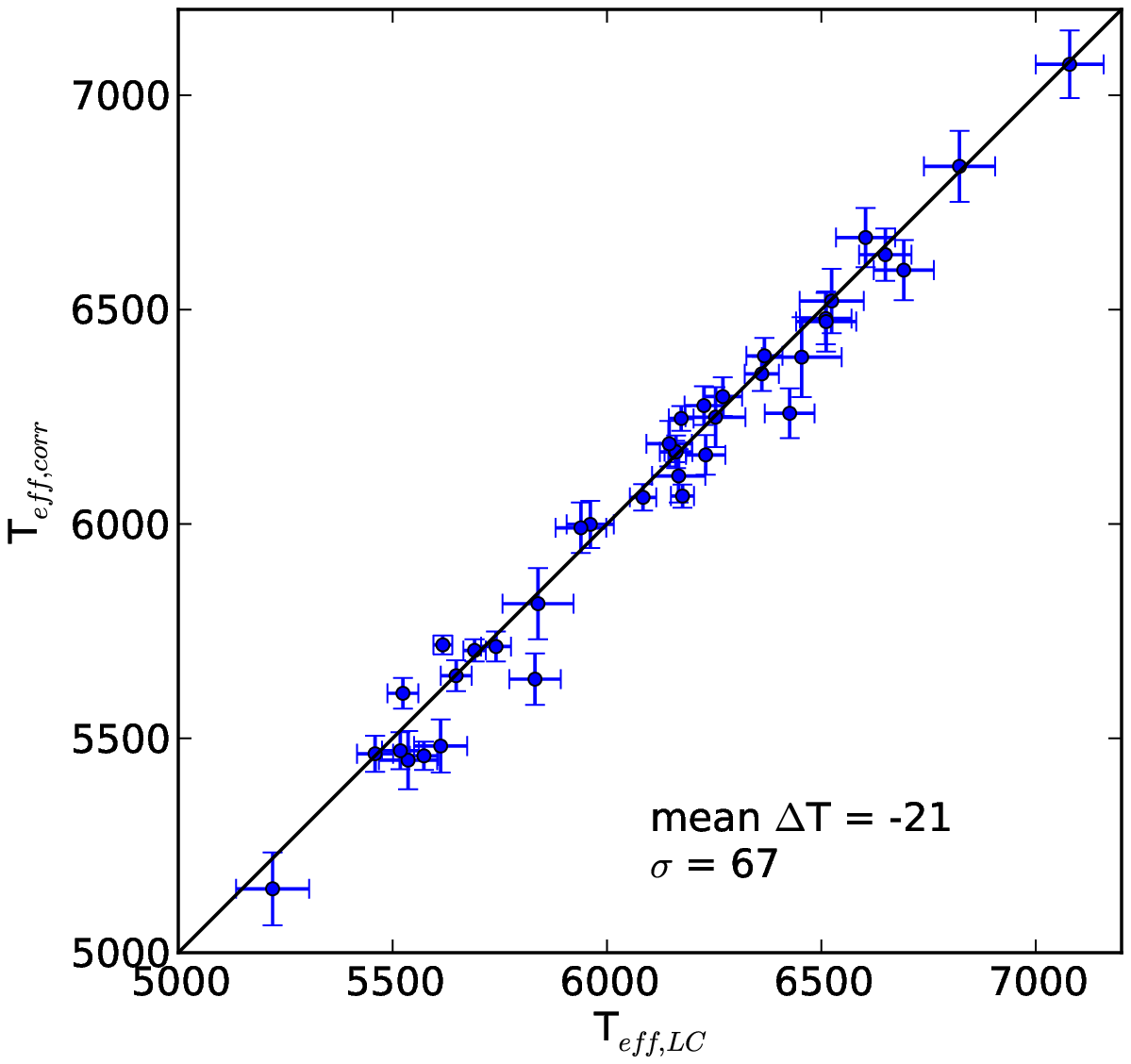}
\includegraphics[width=5.8cm]{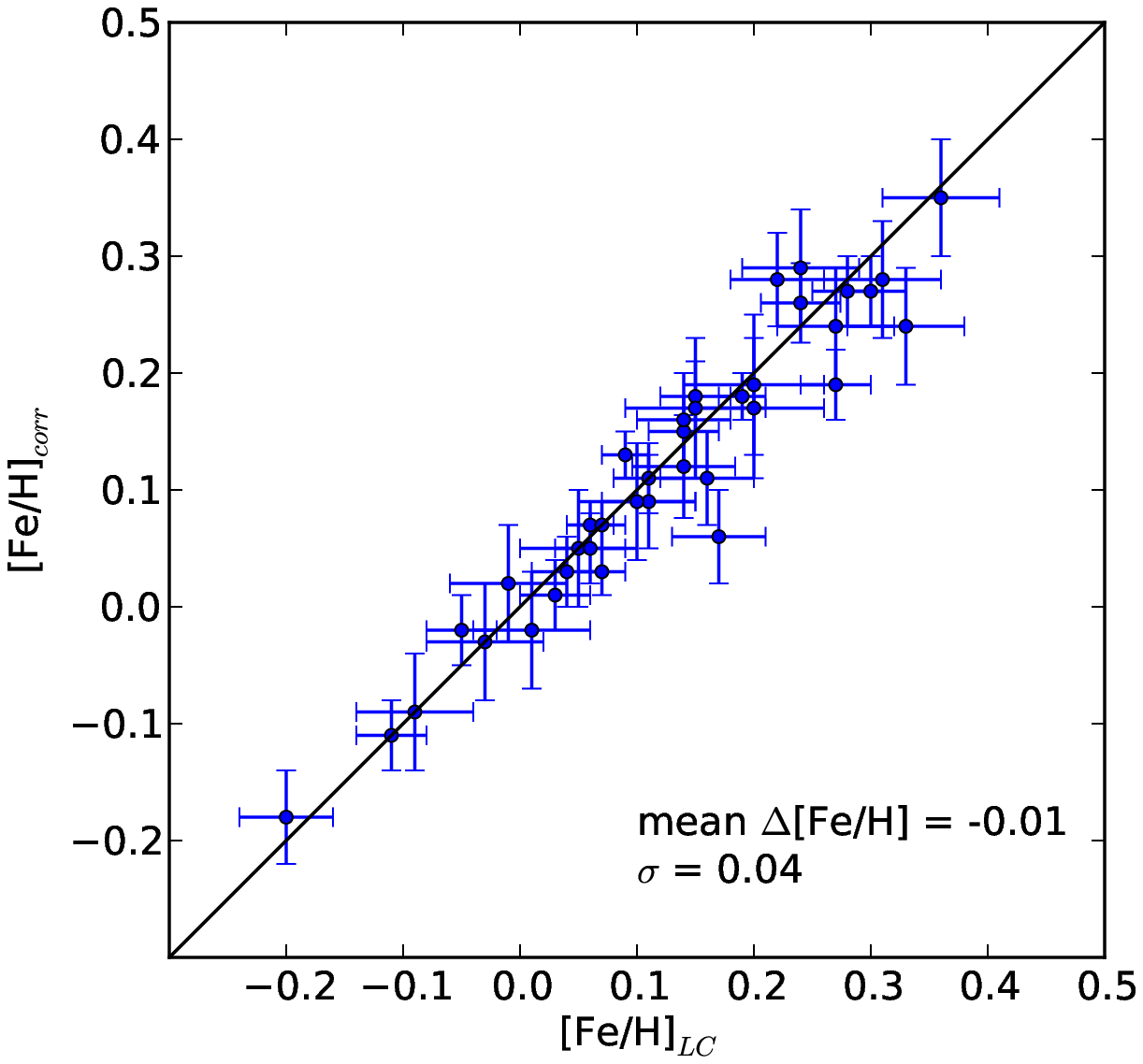}
\includegraphics[width=5.8cm]{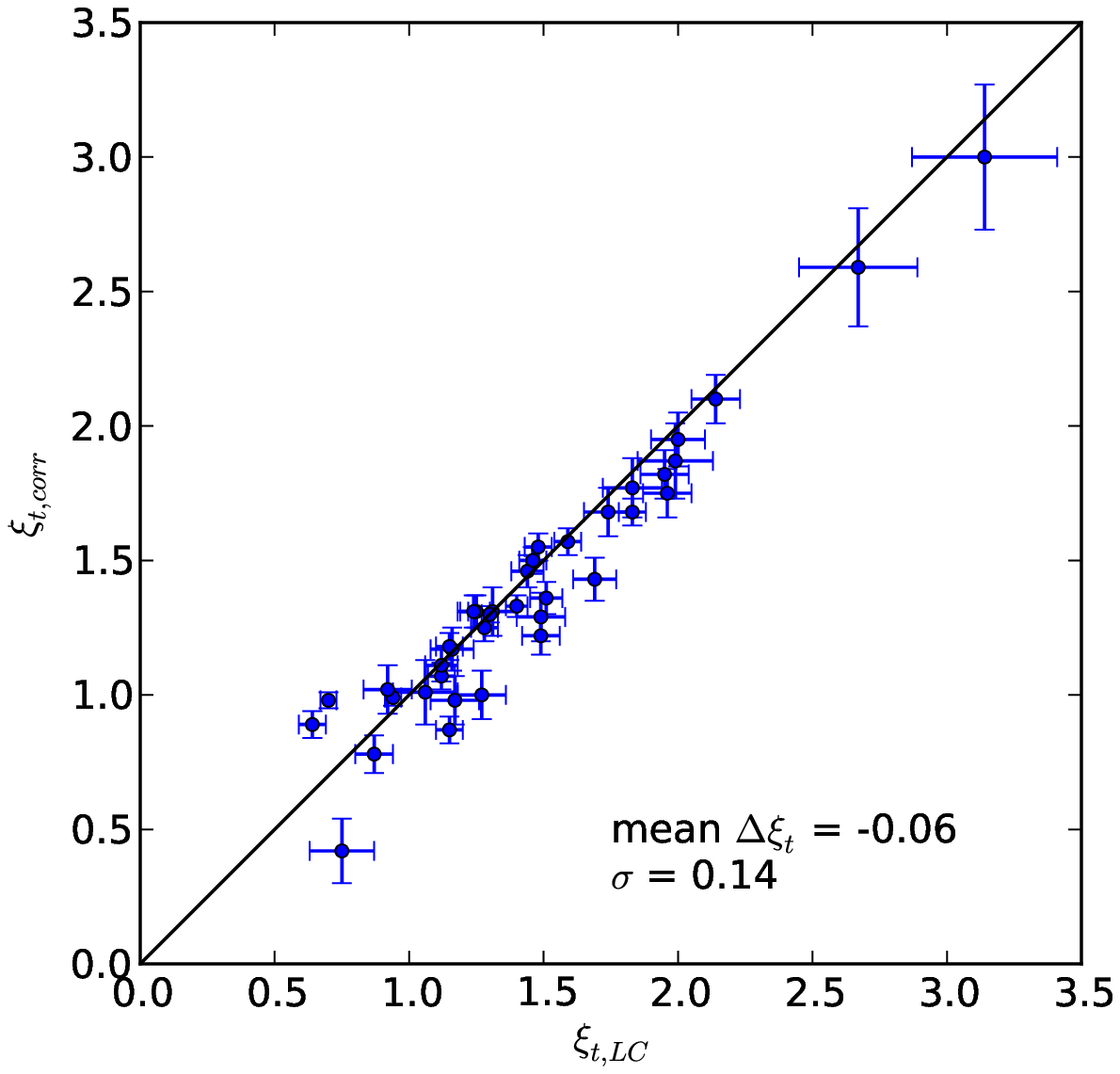}
\caption{Comparisons of the spectroscopic effective temperatures, metallicities, and microturbulences (left to right). In the top panels we compare the unconstrained results (`spec') with the results using a fixed surface gravity from the correction formula (`corr'). In the bottom panels the comparison is shown between the two fixed results ($\log g$ from transit (`LC') and $\log g$ from the formula (`corr')).}
\label{FigFixCorr}
\end{center}
\end{figure*}

We compare the spectroscopic parameters obtained from fixing the surface gravity to the formula corrected value ('corr') with the unconstrained spectroscopic parameters ('spec') and the ones obtained from fixing the surface gravity to the photometric light curve value ('LC'). All parameters compare really well (see Figure \ref{FigFixCorr}), with mean differences of $13$\,K, $0.02$\,dex and $0.02$\,km/s for the effective temperature, metallicity, and microturbulence, respectively for the difference between the corrected values and the spectroscopic values. For the differences between the corrected and the photometric results, we find mean differences of $-21$\,K, $-0.01$\,dex and $-0.06$\,km/s for the effective temperature, metallicity, and microturbulence, respectively. No obvious trends are present. For completeness we calculated the MADs again. For the difference between the corrected values and the spectroscopic values, we find a MAD of $38$\,K, $0.02$\,dex and $0.08$\,km/s for the effective temperature, metallicity, and microturbulence, respectively. The difference between the corrected values and the photometric values gives a MAD of $42$\,K, $0.02$\,dex and $0.08$\,km/s for the effective temperature, metallicity, and microturbulence, respectively. These are thus well within the error bars.

\section{Surface gravity from asteroseismology}\label{Seis}

As \citet{Hub13} showed, the surface gravities obtained through the stellar density from the transit light curve may also be less accurate when the eccentricity or the impact parameter of the transiting planet are under- or overestimated or fixed whilst fitting the light curve. Asteroseismic $\log g$'s on the other hand are more accurate. Although most of the planets from our sample in the previous Section have almost circular orbits, it is still worth, especially since \citet{Hub13} show clear trends, to check if a similar relation can be found to correct spectroscopic surface gravities if one would use asteroseismic surface gravities.

We used a sample compiled from the literature for which the asteroseismic parameters, the maximum frequency $\nu_{max}$ and the large separation $\Delta\nu$, are precisely determined and we have access to high-resolution spectra with moderate to high signal-to-noise. In the end, we have a sample of 86 stars, subsamples of the samples in \citet{Chap14} and \citet{Bru10}. The first work contains asteroseismic data obtained with the Kepler space telescope \citep{Bor09}. The latter compiles a sample of stars analysed with HARPS \citep{Mayor03}.

Spectroscopic parameters for the sample of \citet{Chap14} are gathered from \citet{Mol13}. Their work contains spectroscopic parameters for Kepler targets derived by several methods, one of which is our method with the linelist of \citet{Sou08} as described in Section \ref{Meth}. There are 74 stars in common. The 12 stars from \citet{Bru10} have been spectroscopically analysed with our method either in previous works \citep{San05,Sou08,Tsa13,San13} or in this work.

For stars that were previously not yet analysed by our team, we gathered per star 40 spectra from the HARPS archives (taken from the long asteroseismology series). We shifted them to the reference frame and added them together. Given that these stars are bright, this gives for a high S/N spectrum in the end. We then analysed them following the method described in Section \ref{Meth}. The results are in Table \ref{TabPar}.

The surface gravities of the final sample of 86 stars are then obtained through isochrone fitting using the PARSEC isochrones \citep{Bre12} in the web interface for the Bayesian estimation of stellar parameters\footnote{\url{http://stev.oapd.inaf.it/cgi-bin/param}} \citep[for details, see ][]{Das06}. As input parameters we needed the large separation $\Delta\nu$, the maximum frequency $\nu_{max}$, the effective temperature T$_{\mathrm{eff}}$, and the metallicity [Fe/H]. As Bayesian priors we assumed the lognormal initial mass function from \citet{Chab03} and a constant star formation rate.

\begin{figure}[t!]
\begin{center}
\includegraphics[width=7.0cm]{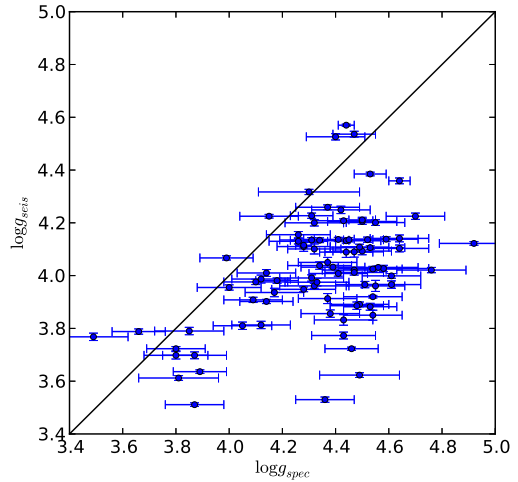}
\caption{Asteroseismic versus spectroscopic surface gravity.}
\label{FigloggSA}
\end{center}
\end{figure}

As expected, the spectroscopic and the asteroseismic surface gravities do not compare well (see Figure \ref{FigloggSA}). As before, we redid, for most of the sample, the same spectroscopic analysis as performed in \citet{ME13c}, but this time we fixed the surface gravity to the asteroseismic value. The results can be found in Table \ref{TabParFix}. 

\begin{figure*}[th!]
\begin{center}
\includegraphics[width=5.8cm]{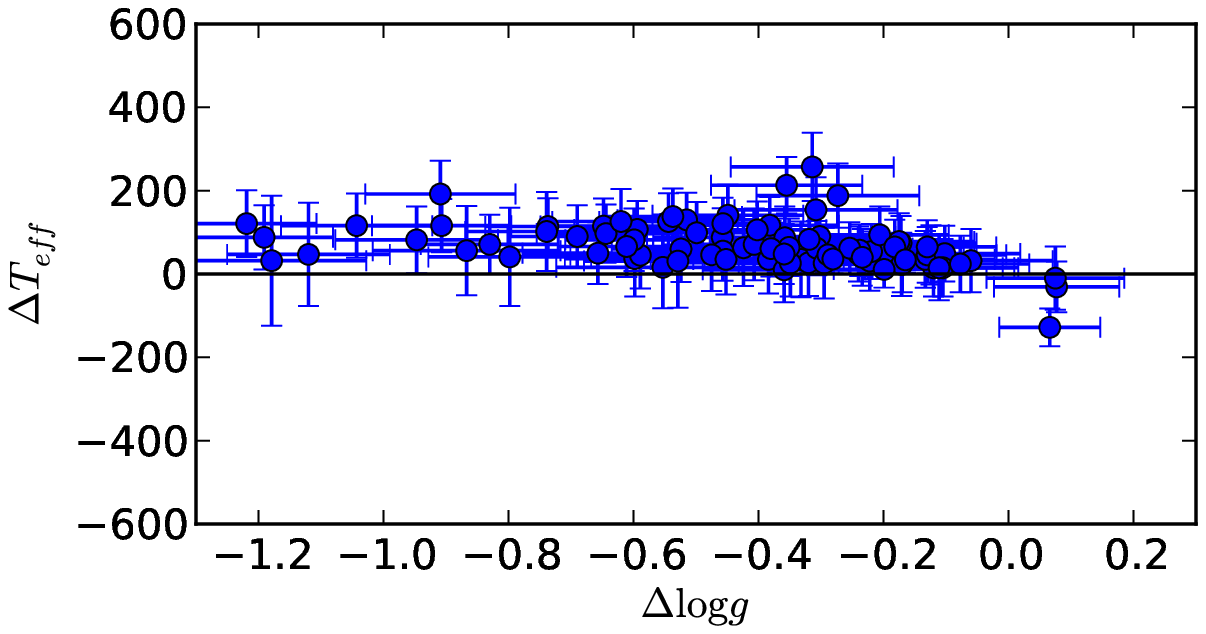}
\includegraphics[width=5.8cm]{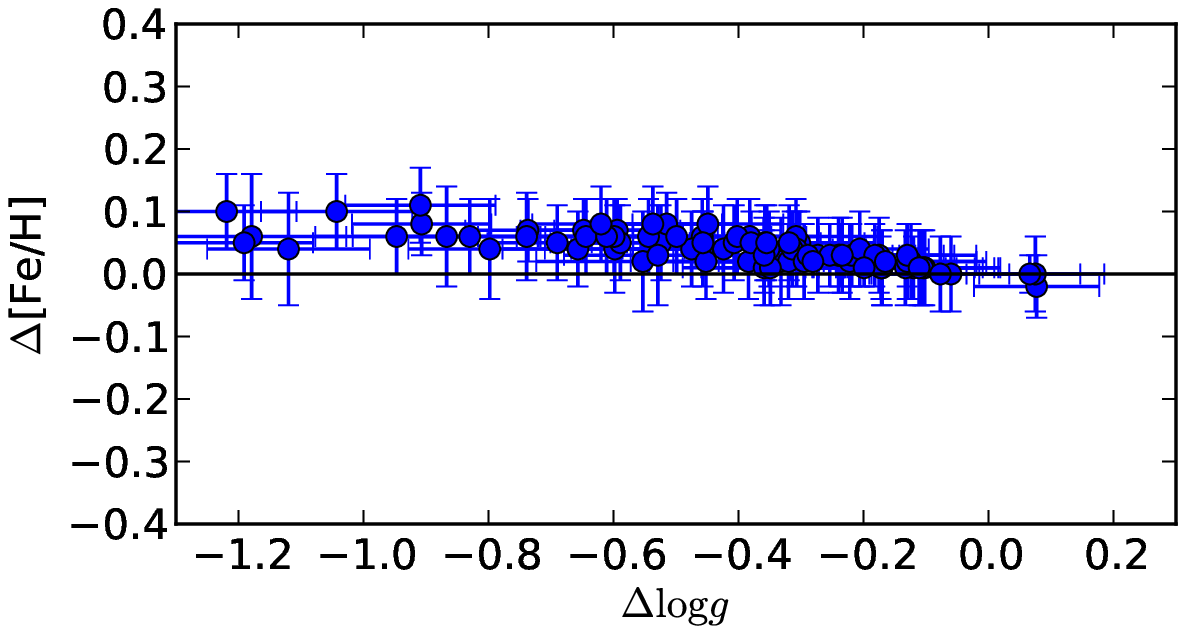}
\includegraphics[width=5.8cm]{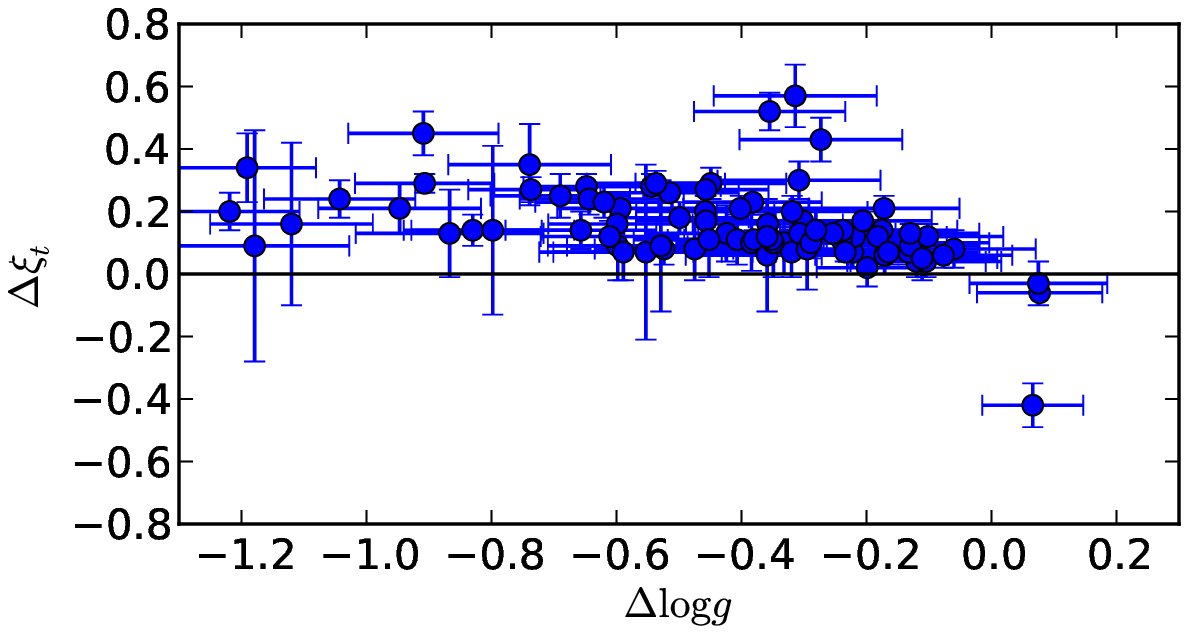}
\caption{Differences of the spectroscopic results (left to right: effective temperature, metallicity, and microturbulence) as a function of the difference in $\log$g (defined as `constrained with asteroseismic $\log g$ - unconstrained').}
\label{FigFA1}
\end{center}
\end{figure*}

We compared the parameters obtained from fixing the surface gravity to the asteroseismic value with the parameters obtained with no constraints on the surface gravity. All parameters compare well, with mean differences of $68$\,K, $0.04$\,dex, and $0.15$\,km/s for the effective temperature, metallicity, and microturbulence, respectively.
In Figure \ref{FigFA1}, the differences in the spectroscopic parameters (defined as `constrained with asteroseismic $\log g$s - unconstrained') are plotted against the difference in surface gravity (defined as `asteroseismic - spectroscopic'). All parameters are slightly anticorrelated with the difference in surface gravity, although most values stay within errorbars. Furthermore, we see the same converging trends as before.

Because of these trends, we again calculated the median absolute deviations (MAD) to quantify the variation. We find that the MADs are $28.5$\,K, $0.02$\,dex, and $0.06$\,km/s for the effective temperature, metallicity, and microturbulence, respectively. Since these values are definitely within the errorbars of the parameters, these trends are thus small enough so that we are again confident that the surface gravity does not have a large effect on the determination of other atmospheric parameters using our method of spectral line analysis and the mentioned linelists. This result confirms the results from Section \ref{Tra}.

\subsection{Correction with temperature}\label{CorrS}

\begin{figure}[t!]
\begin{center}
\includegraphics[width=6.8cm]{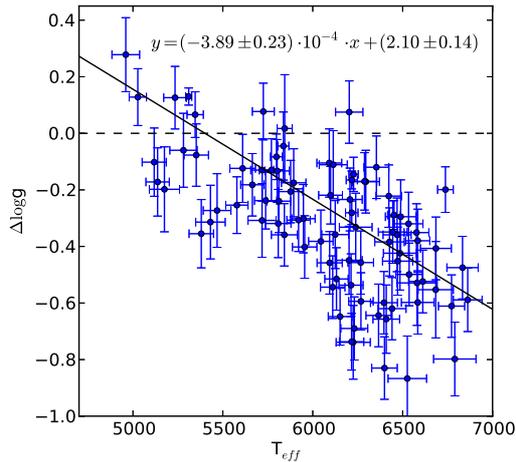}
\caption{Surface gravity difference (`asteroseismic - spectroscopic') versus the (unconstrained) effective temperature. A linear fit is shown with the solid black curve.}
\label{Figlogg2}
\end{center}
\end{figure}

The differences in asteroseismic and spectroscopic surface gravity also seem to depend on the (unconstrained) effective temperature as can be seen in Figure \ref{Figlogg2}, where a decreasing linear trend is again noticeable. The same trend is found for the microturbulence while comparing the $\log$g differences with metallicities reveals no additional trends.

We applied the same procedure as in Section \ref{CorrTr} on the complete sample of 86 stars.
We found the following relation:

\begin{equation}\label{EqFit2}
\log g_{seis} - \log g_{spec} = -3.89\pm0.23\cdot 10^{-4} \cdot T_{\mathrm{eff}} + 2.10 \pm 0.14
\end{equation}

This formula is comparable to the fit presented in Section \ref{CorrTr} for the overlapping temperature range ($5200$\,K till $7000$\,K). This may be somehow surprising since the transit $\log g$ may be less accurate than the asteroseismic one, as showed by \citet{Hub13}. However, we note that in our sample of transiting hosts, most planets have nearly circular orbits which strengthens the accuracy for the derived surface gravity through the transit light curve.

Given the better accuracy of asteroseismic surface gravities as compared to photometric surface gravities, we prefer Equation \ref{EqFit2} to correct for the spectroscopic surface gravity. Since we barely have asteroseismic data for stars cooler than 5200\,K, we cannot guarantee the accuracy of this formula for that temperature range and Equation \ref{EqFit} may thus be preferred for cooler stars.

\section{Summary and discussion}\label{Dis}

In this work we derived spectroscopic parameters (effective temperature, metallicity, surface gravity and microturbulence) for a sample of FGK dwarfs in several ways. First we left the surface gravity free in the spectroscopic analysis as described in Section \ref{Meth} \citep[for the values, see ][and this work]{ME13c,Mol13}. Afterwards, we reran the same analysis whilst fixing the surface gravity to different values:

\begin{itemize}
\item A value obtained through the photometric transit light curve.
\item A value obtained through the large separation and maximum frequency from asteroseismology.
\item A value obtained through an empirical formula, using the effective temperature and the unconstrained surface gravity.
\end{itemize}

We find that, in almost all cases, the resulting stellar atmospheric parameters ($T_{\mathrm{eff}}$, [Fe/H], $\xi_t$) compare well within errorbars although there are slight trends noticable which correlate with the difference in surface gravity. The trends quickly converge and the differences in atmospheric parameters stay stable even for very large differences in surface gravity.

Differences between the constrained and the unconstrained atmospheric parameters can lead to differences in the values for the stellar mass and radius, and thus the planetary mass and radius. On average, the difference for the efffective temperature is about $70$\,K and for the metallicity about $0.04$\,dex. Using these numbers and the calibration formulae from \citet{Tor10}, we find that the resulting stellar mass and stellar radius will, on average, only differ by about $2-3\%$ and $1-1.5\%$, respectively, for FGK dwarfs. This will lead to an average difference of $1.3-2\%$ and $1-1.5\%$ for the planetary mass and radius. These differences are well within the precision that can currently be achieved \citep[e.g.][]{Hub13}.

It seems that the difference between the spectroscopic surface gravities and the photometric or asteroseismic ones is dependent on the effective temperature. By fitting a linear relation to the data, we obtained a correction formula for the surface gravity obtained with our spectroscopic method. Since asteroseismic surface gravities are the most accurate, we recommend to use Equation \ref{EqFit2} rather than Equation \ref{EqFit}. For stars cooler than 5200\,K, we have little asteroseismic data and as such we cannot guarantee the accuracy of the formula for cooler stars. However, since Equation \ref{EqFit} from the photometric $\log g$s is comparable to the one coming from asteroseismic $\log g$s, the former may be used with caution for the cooler stars.

We note that although the surface gravity as calculated through these formulas may be more accurate, it cannot be more precise than the original unconstrained surface gravity, since the error bars of the spectroscopic $\log g$ are factored in when calculating the corrected value for $\log g$. Regardless, the value will definitely be more accurate than the one from the unconstrained MOOG analysis using our proposed linelists. As such the corrected value is better used for calculating other stellar parameters like the stellar mass and radius in case no additional methods can be used to derive the surface gravity, such as a transit light curve or asteroseismology.

\begin{figure}[t!]
\begin{center}
\includegraphics[width=9cm]{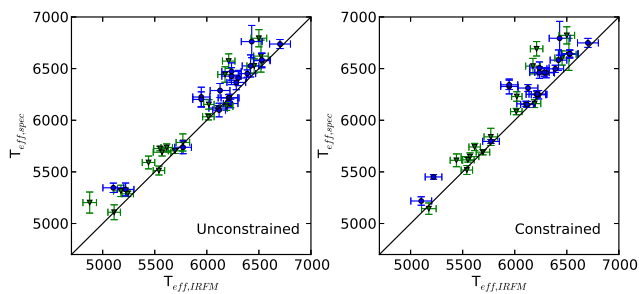}
\caption{Comparisons of the unconstrained (left) and constrained (right) spectroscopic temperatures with literature temperatures obtained through the IRFM method (taken from \citet{Max11b}, represented by green triangles, and \citet{Cas11}, represented by blue circles).}
\label{FigIRFM}
\end{center}
\end{figure}

For the other spectroscopic parameters the question remains whether the original spectroscopic parameters are accurate and can thus be used without performing the spectroscopic analysis again. In the case of the effective temperature, we compared our values with values obtained with the accurate and trusted InfraRed Flux Method (IRFM). We have 19 values from the transit sample from \citet{Max11b} and 21 from \citet{Cas11} of which 2 from the transit sample and 19 from the asteroseismic sample. The comparisons can be seen in Figure \ref{FigIRFM}.

For stars cooler than 6300\,K, the results compare well. We find mean differences of $-75 \pm 100$\,K and $-66\pm74$\,K, for our unconstrained and constrained temperatures, respectively. For the total sample, we find mean differences of $-106\pm122$\,K and $-122\pm138$\,K, respectively. For both the unconstrained and the constrained values, the hotter stars show larger differences, where the spectroscopic temperatures are larger than the ones from the IRFM. This may be an effect of the linelist used for the spectral line analysis. This linelist was calibrated for solar-like stars and the resulting effective temperatures may be overestimated for stars that are much hotter than our Sun \citep[see also][]{Sou11b}.

Given on one hand the marginal difference between comparing the IRFM temperatures with the constrained or the unconstrained temperatures and on the other hand the fact that fixing the surface gravity barely affects the other atmospheric parameters, we can be confident about the results of our unconstrained spectroscopic analysis for the derivation of the effective temperature, metallicity, and microturbulence of FGK dwarfs.

\citet{Tor12} did a similar analysis, but they used an analysis based on synthetic spectra. As already mentioned in Section \ref{Tra}, our results are better constrained than those from an analysis with synthetic spectra and the linelist of \citet{Val05}. They found a linear relation between the temperature and metallicity differences with the surface gravity difference. For surface gravity differences $\Delta \log g \sim 0.5$\,dex, they found differences in temperature of about 350\,K and in metallicity of about 0.20\,dex. With our spectral line analysis method and the carefully selected linelist, we have differences of only 120\,K and 0.05\,dex for temperature and metallicity, respectively.

To conclude, when atmospheric stellar parameters of FGK dwarfs are derived with high-resolution spectroscopy using our ARES+MOOG method, as described in \citet[][and references therein]{San13}, and the linelist of \citet{Sou08} or \citet{Tsa13}, we are confident that the resulting effective temperature, metallicity, and microturbulence are accurate and precise\footnote{We note that the effective temperatures are slightly overestimated for the hotter stars, as mentioned in \citet{Sou11b}.}. The less accurate surface gravity can then easily be corrected using Equation \ref{EqFit2} (or Equation \ref{EqFit} for the coolest stars). This method will always work with high-resolution spectra, even when no other means are available for the determination of surface gravity, like a transit light curve or asteroseismology.

\begin{acknowledgements}
We like to thank the anonymous referee for the fruitful discussion on our paper.
      This work made use of the ESO archive and the Simbad Database. This work was supported by the European Research Council/European Community under the FP7 through Starting Grant agreement number 239953. N.C.S. was supported by FCT through the Investigador FCT contract reference IF/00169/2012 and POPH/FSE (EC) by FEDER funding through the program "Programa Operacional de Factores de Competitividade - COMPETE. V.Zh.A., S.G.S., and I.M.B. acknowledge the support of the Funda\c c\~ao para a Ci\^encia e a Tecnologia (FCT) in the form of grant references SFRH/BPD/70574/2010, SFRH/BPD/47611/2008, and SFRH/BPD/87857/2012. IMB acknowledges support from the EC Project SPACEINN (FP7-SPACE-2012-312844).

\end{acknowledgements}

\bibliographystyle{aa} 
\bibliography{/home/annelies/My_Articles/References.bib}

\begin{thebibliography}{51}
\expandafter\ifx\csname natexlab\endcsname\relax\def\natexlab#1{#1}\fi

\bibitem[{{Adibekyan} {et~al.}(2013{\natexlab{a}}){Adibekyan}, {Figueira},
  {Santos}, {Hakobyan}, {Sousa}, {Pace}, {Delgado Mena}, {Robin}, {Israelian},
  \& {Gonz{\'a}lez Hern{\'a}ndez}}]{Adi13b}
{Adibekyan}, V.~Z., {Figueira}, P., {Santos}, N.~C., {et~al.}
  2013{\natexlab{a}}, \aap, 554, A44

\bibitem[{{Adibekyan} {et~al.}(2013{\natexlab{b}}){Adibekyan}, {Figueira},
  {Santos}, {Mortier}, {Mordasini}, {Delgado Mena}, {Sousa}, {Correia},
  {Israelian}, \& {Oshagh}}]{Adi13}
{Adibekyan}, V.~Z., {Figueira}, P., {Santos}, N.~C., {et~al.}
  2013{\natexlab{b}}, \aap, 560, A51

\bibitem[{{Beaug{\'e}} \& {Nesvorn{\'y}}(2013)}]{Bea13}
{Beaug{\'e}}, C. \& {Nesvorn{\'y}}, D. 2013, \apj, 763, 12

\bibitem[{{Bensby} {et~al.}(2005){Bensby}, {Feltzing}, {Lundstr{\"o}m}, \&
  {Ilyin}}]{Bens05}
{Bensby}, T., {Feltzing}, S., {Lundstr{\"o}m}, I., \& {Ilyin}, I. 2005, \aap,
  433, 185

\bibitem[{{Borucki} {et~al.}(2009){Borucki}, {Koch}, {Jenkins}, {Sasselov},
  {Gilliland}, {Batalha}, {Latham}, {Caldwell}, {Basri}, {Brown},
  {Christensen-Dalsgaard}, {Cochran}, {DeVore}, {Dunham}, {Dupree}, {Gautier},
  {Geary}, {Gould}, {Howell}, {Kjeldsen}, {Lissauer}, {Marcy}, {Meibom},
  {Morrison}, \& {Tarter}}]{Bor09}
{Borucki}, W.~J., {Koch}, D., {Jenkins}, J., {et~al.} 2009, Science, 325, 709

\bibitem[{{Bowler} {et~al.}(2010){Bowler}, {Johnson}, {Marcy}, {Henry}, {Peek},
  {Fischer}, {Clubb}, {Liu}, {Reffert}, {Schwab}, \& {Lowe}}]{Bow10}
{Bowler}, B.~P., {Johnson}, J.~A., {Marcy}, G.~W., {et~al.} 2010, \apj, 709,
  396

\bibitem[{{Bressan} {et~al.}(2012){Bressan}, {Marigo}, {Girardi}, {Salasnich},
  {Dal Cero}, {Rubele}, \& {Nanni}}]{Bre12}
{Bressan}, A., {Marigo}, P., {Girardi}, L., {et~al.} 2012, \mnras, 427, 127

\bibitem[{{Bruntt} {et~al.}(2010){Bruntt}, {Bedding}, {Quirion}, {Lo Curto},
  {Carrier}, {Smalley}, {Dall}, {Arentoft}, {Bazot}, \& {Butler}}]{Bru10}
{Bruntt}, H., {Bedding}, T.~R., {Quirion}, P.-O., {et~al.} 2010, \mnras, 405,
  1907

\bibitem[{{Buchhave} {et~al.}(2014){Buchhave}, {Bizzarro}, {Latham},
  {Sasselov}, {Cochran}, {Endl}, {Isaacson}, {Juncher}, \& {Marcy}}]{Buch14}
{Buchhave}, L.~A., {Bizzarro}, M., {Latham}, D.~W., {et~al.} 2014, \nat, 509,
  593

\bibitem[{{Buchhave} {et~al.}(2012){Buchhave}, {Latham}, {Johansen},
  {Bizzarro}, {Torres}, {Rowe}, {Batalha}, {Borucki}, {Brugamyer}, {Caldwell},
  {Bryson}, {Ciardi}, {Cochran}, {Endl}, {Esquerdo}, {Ford}, {Geary},
  {Gilliland}, {Hansen}, {Isaacson}, {Laird}, {Lucas}, {Marcy}, {Morse},
  {Robertson}, {Shporer}, {Stefanik}, {Still}, \& {Quinn}}]{Buch12}
{Buchhave}, L.~A., {Latham}, D.~W., {Johansen}, A., {et~al.} 2012, \nat, 486,
  375

\bibitem[{{Butler} {et~al.}(2006){Butler}, {Johnson}, {Marcy}, {Wright},
  {Vogt}, \& {Fischer}}]{But06}
{Butler}, R.~P., {Johnson}, J.~A., {Marcy}, G.~W., {et~al.} 2006, \pasp, 118,
  1685

\bibitem[{{Casagrande} {et~al.}(2011){Casagrande}, {Sch{\"o}nrich}, {Asplund},
  {Cassisi}, {Ram{\'{\i}}rez}, {Mel{\'e}ndez}, {Bensby}, \& {Feltzing}}]{Cas11}
{Casagrande}, L., {Sch{\"o}nrich}, R., {Asplund}, M., {et~al.} 2011, \aap, 530,
  A138

\bibitem[{{Chabrier}(2003)}]{Chab03}
{Chabrier}, G. 2003, \pasp, 115, 763

\bibitem[{{Chaplin} {et~al.}(2014){Chaplin}, {Basu}, {Huber}, {Serenelli},
  {Casagrande}, {Silva Aguirre}, {Ball}, {Creevey}, {Gizon}, {Handberg},
  {Karoff}, {Lutz}, {Marques}, {Miglio}, {Stello}, {Suran}, {Pricopi},
  {Metcalfe}, {Monteiro}, {Molenda-{\.Z}akowicz}, {Appourchaux},
  {Christensen-Dalsgaard}, {Elsworth}, {Garc{\'{\i}}a}, {Houdek}, {Kjeldsen},
  {Bonanno}, {Campante}, {Corsaro}, {Gaulme}, {Hekker}, {Mathur}, {Mosser},
  {R{\'e}gulo}, \& {Salabert}}]{Chap14}
{Chaplin}, W.~J., {Basu}, S., {Huber}, D., {et~al.} 2014, \apjs, 210, 1

\bibitem[{{da Silva} {et~al.}(2006){da Silva}, {Girardi}, {Pasquini},
  {Setiawan}, {von der L{\"u}he}, {de Medeiros}, {Hatzes}, {D{\"o}llinger}, \&
  {Weiss}}]{Das06}
{da Silva}, L., {Girardi}, L., {Pasquini}, L., {et~al.} 2006, \aap, 458, 609

\bibitem[{{Dumusque} {et~al.}(2014){Dumusque}, {Bonomo}, {Haywood},
  {Malavolta}, {S{\'e}gransan}, {Buchhave}, {Collier Cameron}, {Latham},
  {Molinari}, {Pepe}, {Udry}, {Charbonneau}, {Cosentino}, {Dressing},
  {Figueira}, {Fiorenzano}, {Gettel}, {Harutyunyan}, {Horne}, {Lopez-Morales},
  {Lovis}, {Mayor}, {Micela}, {Motalebi}, {Nascimbeni}, {Phillips}, {Piotto},
  {Pollacco}, {Queloz}, {Rice}, {Sasselov}, {Sozzetti}, {Szentgyorgyi}, \&
  {Watson}}]{Dum14}
{Dumusque}, X., {Bonomo}, A.~S., {Haywood}, R.~D., {et~al.} 2014, \apj, 789,
  154

\bibitem[{{Edvardsson} {et~al.}(1993){Edvardsson}, {Andersen}, {Gustafsson},
  {Lambert}, {Nissen}, \& {Tomkin}}]{Edv93}
{Edvardsson}, B., {Andersen}, J., {Gustafsson}, B., {et~al.} 1993, \aap, 275,
  101

\bibitem[{{Haywood}(2008)}]{Hay08}
{Haywood}, M. 2008, \mnras, 388, 1175

\bibitem[{{Huber} {et~al.}(2013){Huber}, {Chaplin}, {Christensen-Dalsgaard},
  {Gilliland}, {Kjeldsen}, {Buchhave}, {Fischer}, {Lissauer}, {Rowe},
  {Sanchis-Ojeda}, {Basu}, {Handberg}, {Hekker}, {Howard}, {Isaacson},
  {Karoff}, {Latham}, {Lund}, {Lundkvist}, {Marcy}, {Miglio}, {Silva Aguirre},
  {Stello}, {Arentoft}, {Barclay}, {Bedding}, {Burke}, {Christiansen},
  {Elsworth}, {Haas}, {Kawaler}, {Metcalfe}, {Mullally}, \& {Thompson}}]{Hub13}
{Huber}, D., {Chaplin}, W.~J., {Christensen-Dalsgaard}, J., {et~al.} 2013,
  \apj, 767, 127

\bibitem[{{Johnson} {et~al.}(2010){Johnson}, {Aller}, {Howard}, \&
  {Crepp}}]{John10}
{Johnson}, J.~A., {Aller}, K.~M., {Howard}, A.~W., \& {Crepp}, J.~R. 2010,
  \pasp, 122, 905

\bibitem[{{Kurucz}(1993)}]{Kur93}
{Kurucz}, R. 1993, ATLAS9 Stellar Atmosphere Programs and 2 km/s grid.~Kurucz
  CD-ROM No.~13.~ Cambridge, Mass.: Smithsonian Astrophysical Observatory,
  1993., 13

\bibitem[{{Marcy} {et~al.}(2014){Marcy}, {Isaacson}, {Howard}, {Rowe},
  {Jenkins}, {Bryson}, {Latham}, {Howell}, {Gautier}, {Batalha}, {Rogers},
  {Ciardi}, {Fischer}, {Gilliland}, {Kjeldsen}, {Christensen-Dalsgaard},
  {Huber}, {Chaplin}, {Basu}, {Buchhave}, {Quinn}, {Borucki}, {Koch}, {Hunter},
  {Caldwell}, {Van Cleve}, {Kolbl}, {Weiss}, {Petigura}, {Seager}, {Morton},
  {Johnson}, {Ballard}, {Burke}, {Cochran}, {Endl}, {MacQueen}, {Everett},
  {Lissauer}, {Ford}, {Torres}, {Fressin}, {Brown}, {Steffen}, {Charbonneau},
  {Basri}, {Sasselov}, {Winn}, {Sanchis-Ojeda}, {Christiansen}, {Adams},
  {Henze}, {Dupree}, {Fabrycky}, {Fortney}, {Tarter}, {Holman}, {Tenenbaum},
  {Shporer}, {Lucas}, {Welsh}, {Orosz}, {Bedding}, {Campante}, {Davies},
  {Elsworth}, {Handberg}, {Hekker}, {Karoff}, {Kawaler}, {Lund}, {Lundkvist},
  {Metcalfe}, {Miglio}, {Silva Aguirre}, {Stello}, {White}, {Boss}, {Devore},
  {Gould}, {Prsa}, {Agol}, {Barclay}, {Coughlin}, {Brugamyer}, {Mullally},
  {Quintana}, {Still}, {Thompson}, {Morrison}, {Twicken}, {D{\'e}sert},
  {Carter}, {Crepp}, {H{\'e}brard}, {Santerne}, {Moutou}, {Sobeck}, {Hudgins},
  {Haas}, {Robertson}, {Lillo-Box}, \& {Barrado}}]{Marcy14}
{Marcy}, G.~W., {Isaacson}, H., {Howard}, A.~W., {et~al.} 2014, \apjs, 210, 20

\bibitem[{{Maxted} {et~al.}(2011){Maxted}, {Koen}, \& {Smalley}}]{Max11b}
{Maxted}, P.~F.~L., {Koen}, C., \& {Smalley}, B. 2011, \mnras, 418, 1039

\bibitem[{{Mayor} {et~al.}(2011){Mayor}, {Marmier}, {Lovis}, {Udry},
  {S{\'e}gransan}, {Pepe}, {Benz}, {Bertaux}, {Bouchy}, {Dumusque}, {Lo Curto},
  {Mordasini}, {Queloz}, \& {Santos}}]{Mayor11}
{Mayor}, M., {Marmier}, M., {Lovis}, C., {et~al.} 2011, ArXiv e-prints,
  1109.2497

\bibitem[{{Mayor} {et~al.}(2003){Mayor}, {Pepe}, {Queloz}, {Bouchy},
  {Rupprecht}, {Lo Curto}, {Avila}, {Benz}, {Bertaux}, {Bonfils}, {Dall},
  {Dekker}, {Delabre}, {Eckert}, {Fleury}, {Gilliotte}, {Gojak}, {Guzman},
  {Kohler}, {Lizon}, {Longinotti}, {Lovis}, {Megevand}, {Pasquini}, {Reyes},
  {Sivan}, {Sosnowska}, {Soto}, {Udry}, {van Kesteren}, {Weber}, \&
  {Weilenmann}}]{Mayor03}
{Mayor}, M., {Pepe}, F., {Queloz}, D., {et~al.} 2003, The Messenger, 114, 20

\bibitem[{{McWilliam} {et~al.}(2008){McWilliam}, {Matteucci}, {Ballero},
  {Rich}, {Fulbright}, \& {Cescutti}}]{McW08}
{McWilliam}, A., {Matteucci}, F., {Ballero}, S., {et~al.} 2008, \aj, 136, 367

\bibitem[{{Minchev} {et~al.}(2013){Minchev}, {Chiappini}, \& {Martig}}]{Min13}
{Minchev}, I., {Chiappini}, C., \& {Martig}, M. 2013, \aap, 558, A9

\bibitem[{{Molenda-{\.Z}akowicz} {et~al.}(2013){Molenda-{\.Z}akowicz}, {Sousa},
  {Frasca}, {Uytterhoeven}, {Briquet}, {Van Winckel}, {Drobek}, {Niemczura},
  {Lampens}, {Lykke}, {Bloemen}, {Gameiro}, {Jean}, {Volpi}, {Gorlova},
  {Mortier}, {Tsantaki}, \& {Raskin}}]{Mol13}
{Molenda-{\.Z}akowicz}, J., {Sousa}, S.~G., {Frasca}, A., {et~al.} 2013,
  \mnras, 434, 1422

\bibitem[{{Mordasini} {et~al.}(2012){Mordasini}, {Alibert}, {Benz}, {Klahr}, \&
  {Henning}}]{Mor12}
{Mordasini}, C., {Alibert}, Y., {Benz}, W., {Klahr}, H., \& {Henning}, T. 2012,
  \aap, 541, A97

\bibitem[{{Mortier} {et~al.}(2013{\natexlab{a}}){Mortier}, {Santos}, {Sousa},
  {Israelian}, {Mayor}, \& {Udry}}]{ME13}
{Mortier}, A., {Santos}, N.~C., {Sousa}, S., {et~al.} 2013{\natexlab{a}}, \aap,
  551, A112

\bibitem[{{Mortier} {et~al.}(2013{\natexlab{b}}){Mortier}, {Santos}, {Sousa},
  {Adibekyan}, {Delgado Mena}, {Tsantaki}, {Israelian}, \& {Mayor}}]{ME13b}
{Mortier}, A., {Santos}, N.~C., {Sousa}, S.~G., {et~al.} 2013{\natexlab{b}},
  \aap, 557, A70

\bibitem[{{Mortier} {et~al.}(2013{\natexlab{c}}){Mortier}, {Santos}, {Sousa},
  {Fernandes}, {Adibekyan}, {Delgado Mena}, {Montalto}, \& {Israelian}}]{ME13c}
{Mortier}, A., {Santos}, N.~C., {Sousa}, S.~G., {et~al.} 2013{\natexlab{c}},
  \aap, 558, A106

\bibitem[{Press {et~al.}(1992)Press, Teukolsky, Vetterling, \&
  Flannery}]{Pre92C}
Press, W.~H., Teukolsky, S.~A., Vetterling, W.~T., \& Flannery, B.~P. 1992,
  Numerical Recipes in C (2Nd Ed.): The Art of Scientific Computing (New York,
  NY, USA: Cambridge University Press)

\bibitem[{{Press} {et~al.}(1992){Press}, {Teukolsky}, {Vetterling}, \&
  {Flannery}}]{Pre92}
{Press}, W.~H., {Teukolsky}, S.~A., {Vetterling}, W.~T., \& {Flannery}, B.~P.
  1992, {Numerical recipes in FORTRAN. The art of scientific computing}

\bibitem[{{Ram{\'{\i}}rez} {et~al.}(2013){Ram{\'{\i}}rez}, {Allende Prieto}, \&
  {Lambert}}]{Ram13}
{Ram{\'{\i}}rez}, I., {Allende Prieto}, C., \& {Lambert}, D.~L. 2013, \apj,
  764, 78

\bibitem[{{Recio-Blanco} {et~al.}(2006){Recio-Blanco}, {Bijaoui}, \& {de
  Laverny}}]{Rec06}
{Recio-Blanco}, A., {Bijaoui}, A., \& {de Laverny}, P. 2006, \mnras, 370, 141

\bibitem[{{Santos} {et~al.}(2004){Santos}, {Israelian}, \& {Mayor}}]{San04}
{Santos}, N.~C., {Israelian}, G., \& {Mayor}, M. 2004, \aap, 415, 1153

\bibitem[{{Santos} {et~al.}(2005){Santos}, {Israelian}, {Mayor}, {Bento},
  {Almeida}, {Sousa}, \& {Ecuvillon}}]{San05}
{Santos}, N.~C., {Israelian}, G., {Mayor}, M., {et~al.} 2005, \aap, 437, 1127

\bibitem[{{Santos} {et~al.}(2013){Santos}, {Sousa}, {Mortier}, {Neves},
  {Adibekyan}, {Tsantaki}, {Delgado Mena}, {Bonfils}, {Israelian}, {Mayor}, \&
  {Udry}}]{San13}
{Santos}, N.~C., {Sousa}, S.~G., {Mortier}, A., {et~al.} 2013, \aap, 556, A150

\bibitem[{{Sneden}(1973)}]{Sne73}
{Sneden}, C.~A. 1973, PhD thesis, The University of Texas at Austin.

\bibitem[{{Sousa} {et~al.}(2007){Sousa}, {Santos}, {Israelian}, {Mayor}, \&
  {Monteiro}}]{Sou07}
{Sousa}, S.~G., {Santos}, N.~C., {Israelian}, G., {Mayor}, M., \& {Monteiro},
  M.~J.~P.~F.~G. 2007, \aap, 469, 783

\bibitem[{{Sousa} {et~al.}(2011){Sousa}, {Santos}, {Israelian}, {Mayor}, \&
  {Udry}}]{Sou11b}
{Sousa}, S.~G., {Santos}, N.~C., {Israelian}, G., {Mayor}, M., \& {Udry}, S.
  2011, \aap, 533, A141+

\bibitem[{{Sousa} {et~al.}(2008){Sousa}, {Santos}, {Mayor}, {Udry},
  {Casagrande}, {Israelian}, {Pepe}, {Queloz}, \& {Monteiro}}]{Sou08}
{Sousa}, S.~G., {Santos}, N.~C., {Mayor}, M., {et~al.} 2008, \aap, 487, 373

\bibitem[{{Torres} {et~al.}(2010){Torres}, {Andersen}, \&
  {Gim{\'e}nez}}]{Tor10}
{Torres}, G., {Andersen}, J., \& {Gim{\'e}nez}, A. 2010, \aapr, 18, 67

\bibitem[{{Torres} {et~al.}(2012){Torres}, {Fischer}, {Sozzetti}, {Buchhave},
  {Winn}, {Holman}, \& {Carter}}]{Tor12}
{Torres}, G., {Fischer}, D.~A., {Sozzetti}, A., {et~al.} 2012, \apj, 757, 161

\bibitem[{{Torres} {et~al.}(2008){Torres}, {Winn}, \& {Holman}}]{Tor08}
{Torres}, G., {Winn}, J.~N., \& {Holman}, M.~J. 2008, \apj, 677, 1324

\bibitem[{{Tsantaki} {et~al.}(2013){Tsantaki}, {Sousa}, {Adibekyan}, {Santos},
  {Mortier}, \& {Israelian}}]{Tsa13}
{Tsantaki}, M., {Sousa}, S.~G., {Adibekyan}, V.~Z., {et~al.} 2013, \aap, 555,
  A150

\bibitem[{{Udry} \& {Santos}(2007)}]{Udry07}
{Udry}, S. \& {Santos}, N.~C. 2007, \araa, 45, 397

\bibitem[{{Valenti} \& {Fischer}(2005)}]{Val05}
{Valenti}, J.~A. \& {Fischer}, D.~A. 2005, \apjs, 159, 141

\bibitem[{{Valenti} \& {Piskunov}(1996)}]{Val96}
{Valenti}, J.~A. \& {Piskunov}, N. 1996, \aaps, 118, 595

\bibitem[{{Winn}(2011)}]{Winn11}
{Winn}, J.~N. 2011, {Exoplanet Transits and Occultations}, ed. S.~{Seager},
  55--77

\end{thebibliography}

\end{document}